\documentclass[reprint,twocolumn]{revtex4-2}
\usepackage{amsfonts}
\usepackage{amsmath}
\usepackage{amssymb}
\usepackage{charter}
\usepackage{graphicx}
\usepackage{float}
\usepackage{amsmath}
\usepackage{amssymb}
\usepackage{charter}
\usepackage{graphicx}
\usepackage{subfigure}
\usepackage{graphicx}
\usepackage{epstopdf}
\usepackage{color}
\usepackage{tocvsec2}
\usepackage{enumerate}
\usepackage{graphicx}
\usepackage{dcolumn}
\usepackage{bm}

\begin{document}

\title{Phase sensitivity via photon-subtraction operations inside
Mach-Zehnder interferometer}
\author{Qisi Zhou$^{1}$}
\author{Qinqian Kang$^{1,2}$}
\author{Tao Jiang$^{1}$}
\author{Zekun Zhao$^{1}$}
\author{Teng Zhao$^{1,3}$}
\author{Cunjin Liu$^{1,*}$}
\author{Liyun Hu$^{1,3,}$}
\thanks{Corresponding authors: lcjwelldone@126.com, hlyun@jxnu.edu.cn}
\affiliation{$^{{\small 1}}$\textit{Center for Quantum Science and Technology, Jiangxi
Normal University, Nanchang 330022, China}\\
$^{{\small 2}}$\textit{Department of Physics, Jiangxi Normal University
Science and Technology College, Nanchang 330022, China}\\
$^{{\small 3}}$\textit{Institute for Military-Civilian Integration of Jiangxi Province,
 Nanchang 330200, China}}

\begin{abstract}
Based on the conventional Mach-Zehnder interferometer, we propose a
metrological scheme to improve phase sensitivity. In this scheme, we use a
coherent state and a squeezed vacuum state as input states, employ
multi-photon-subtraction operations and make intensity-detection or
homodyne-detection. We study phase sensitivity, quantum Fisher information
and quantum\ Cram\'{e}r-Rao bound under both ideal and lossy conditions. The
results indicate that choosing an appropriate detection method and photon
subtraction scheme can significantly enhance the phase sensitivity and
robustness against photon losses. Even under lossy conditions, the
multi-photon subtraction schemes can surpass the standard quantum limit.
Notably, the homodyne detection method can even break through the Heisenberg
limit. Moreover, increasing the number of photon-subtracted can enhance both
phase sensitivity and quantum Fisher information. This research highlights
the significant value of this scheme in quantum precision measurement.

\textbf{PACS: }03.67.-a, 05.30.-d, 42.50,Dv, 03.65.Wj
\end{abstract}

\maketitle

\section{Introduction}

Quantum metrology, an emerging discipline that integrates quantum mechanics
with statistical methods (metrology), is attracting increasing interest \cite%
{02,03,04,05,07}. It shows remarkable potential in a wide range of
applications, including quantum lithography \cite{3,4}, quantum imaging \cite%
{5,6}, atomic clocks \cite{7,8}, gravitational wave measurements \cite{9,10}%
, optical gyroscopes and so on \cite{12,13}. The primary research is quantum
precision measurement, which is object to leverage quantum techniques to
achieve maximal measurement accuracy, specifically enhancing estimation
accuracy for the parameters being assessed. In precision measurements,
achieving highly sensitive phase estimation is critical to progress \cite%
{a1,a2,a3,a4}. Classically, the phase sensitivity of a linear interferometer
with a single-mode coherent state as input is bounded by the standard
quantum limit (SQL), which is given by $1/\sqrt{N}$, where $N$ is the
average photon number sensitive to the phase \cite{15}. Moreover, there
exists a higher limit - the Heisenberg limit (HL) $1/N$. By employing
quantum resources and quantum technologies, the phase sensitivity can break
through SQL and even reach the HL. This quantum advantage highlights the
potential of quantum techniques in precision measurements \cite{15}.

In recent decades, numerous researchers have proposed various enhancement
schemes to achieve higher phase sensitivity. These approaches can be
categorized into three main strategies: (i) employing non-classical states,
such as squeezed states and entangled states, as inputs for the
interferometer; (ii) changing the interferometer or incorporating additional
operations within it; (iii) selecting a well detection method at the output.
As early as 1981, Caves showed that squeezed light can improve the phase
sensitivity of Mach-Zehnder interferometer (MZI) below the SQL \cite{c1}. In
order to exceed SQL and achieve higher precision, various quantum sources
have been investigated, including entangled coherent states \cite{03,c3},
twin Fock states \cite{04,05}, NOON states \cite{c6,c7}, two-mode squeezed
vacuum states \cite{c8}, etc. Some not only exceed the SQL but also even
reach the HL. Another promising approach to enhancing phase sensitivity
involves modifying the structural design of the conventional MZI. In 1986,
Yurke \textit{et al}. first proposed the SU(1,1) interferometer by replacing
beam splitters (BSs) with optical parametric amplifiers (OPAs) \cite{c10}.
Additional configurations involve substituting the linear phase shifter with
a Kerr nonlinear phase shifter \cite{c101,c102} and using an
adjustable-ratio BS instead of a 50:50 BS \cite{c103,c104}.

In addition, non-Gaussian operations offer significant advantages in
metrology and quantum computing. Numerous protocols using non-Gaussian
operations (such as photon addition, photon subtraction, photon catalysis,
and number-conserving) have been studied \cite{e1,e2,e3,e4,e6,e7,e8,e10,e11}%
. Theoretically, these operations effectively enhance the non-classicality
and entanglement of quantum states. They are experimentally feasible \cite%
{d01,d02,d03}. Therefore, non-Gaussian operations are commonly used to
prepare non-classical states to improve phase accuracy. For example, Verma
\textit{et al}. have studied the generation of non-Gaussian squeezed vacuum
states under realistic conditions and their improvement of the phase
sensitivity of MZI \cite{d21}. And Kumar \textit{et al}. have studied the
use of non-Gaussian two-mode squeezed thermal input states to enhance the
phase estimation of MZI \cite{d22}. Besides, it has been demonstrated in
related studies that implementing non-Gaussian operations inside the SU(1,1)
interferometer can effectively enhance phase sensitivity and mitigate the
impact of internal photon losses. For instance, Xu \textit{et al}.
investigated the phase sensitivity enhancement achieved via photon addition
within an SU(1,1) interferometer by using intensity detection. Their study
demonstrates that performing photon addition operations internally provides
superior results compared to those at the input \cite{d23}. Kang \textit{et
al}. investigated the phase sensitivity enhancement achieved via photon
subtraction within the SU(1,1) interferometer by using a homodyne detection
scheme \cite{d24}. However, there are few relevant articles that consider
non-Gaussian operations within the MZI, which is one of the most widely used
interferometers in phase estimation. The impact of performing non-Gaussian
operations inside MZI on phase sensitivity remains unclear.\ Therefore, we
propose multi-photon subtraction schemes (multi-PSSs) by performing photon
subtraction operations within the MZI, including photon subtraction acting
on\ mode $a$,\ mode $b$, and both modes.

It is known that the detection method also influences the accuracy of phase
measurements. While numerous articles often select a specific measurement
method for discussion, there exists a variety of detection methods to choose
from, and each method has its own advantages and disadvantages. The choice
of detection method is closely related to the system studied. It is also an
intrinsic factor that would affect the results of the non-Gaussian operation
on phase sensitivity. However, in Refs. \cite{d23} and \cite{d24}, the
studies conducted by Xu \textit{et al}. and Kang \textit{et al}. were based
on a specific detection scheme and did not involve optimizing the choice of
detection method. In this paper, we shall study the phase sensitivity of MZI
with multi-PSSs under different detection methods to identify appropriate
detection methods and the multi-PSSs. Based on the appropriate selection, we
will further analyze the results on phase sensitivity in both ideal and
lossy cases.

The paper is structured as follows. In Sec. II, we introduce the mode of
multi-PSSs. In Sec. III, we investigate the phase sensitivity for multi-PSSs
under various detection methods in both ideal and internal photon losses
scenarios. In Sec. IV, we investigate the impact of multi-PSSs on QFI and
QCRB. Finally, we give the conclusion in Sec. V.

\section{Proposed scheme}

In this section, we firstly introduce the standard MZI without multi-PSSs,
which comprises two BSs and a linear phase shifter. The first BS is
characterized by operator $\hat{B}_{1}=\exp [-i\pi (\hat{a}^{\dagger }\hat{b}%
+\hat{a}\hat{b}^{\dagger })/4]$, where $\hat{a}$ ($\hat{b}$), $\hat{a}%
^{\dagger }$ ($\hat{b}^{\dagger }$) represent the photon annihilation and
photon creation operators, respectively. Following the first BS, mode $a$
undergoes a phase shift process $\hat{U}_{\phi }=\exp [i\phi (\hat{a}%
^{\dagger }\hat{a})]$, while mode $b$ remains unchanged. Subsequently, the
two beams are coupled in the second BS with the operator $\hat{B}_{2}=\exp
[i\pi (\hat{a}^{\dagger }\hat{b}+\hat{a}\hat{b}^{\dagger })/4]$. For
arbitrary given input states $\left \vert \psi \right \rangle
_{in}=\left
\vert \varphi \right \rangle _{a}\otimes \left \vert \varphi
\right \rangle _{b}$, the output state of a lossless standard MZI can be
expressed as $\left \vert \psi \right \rangle _{out}=\hat{B}_{2}\hat{U}%
_{\phi }\hat{B}_{1}\left \vert \psi \right \rangle _{in}$.

The above model of the MZI is based on the ideal case, without the
consideration of the system loss. However, losses is inevitable in practical
experimental situations. In this paper, we only consider the internal
phonton losses inside the MZI. Theoretically, the phonton losses can be
simulated by fictitious BSs, the operator of which is represented as $\hat{B}%
_{L}=\hat{B}_{La}\otimes \hat{B}_{Lb}$, with $\hat{B}_{La}=\exp [\theta _{a}(%
\hat{a}^{^{\dagger }}\hat{a}_{v}-\hat{a}\hat{a}_{v}^{^{\dagger }})]$ and $%
\hat{B}_{Lb}=\exp [\theta _{b}(\hat{b}^{\dagger }\hat{b}_{v}-\hat{b}\hat{b}%
_{v}^{\dagger })]$, where $\hat{a}_{v}$ and $\hat{b}_{v}$ represent vacuum
modes. Here, $T_{k}$ ($k=a,b$) denotes the transmissivity of the fictitious
BSs, associated with $\theta _{k}$ through $T_{k}=\cos ^{2}\theta _{k}\in %
\left[ 0,1\right] $. The lossless case corresponds to that $T_{k}=1$ \cite%
{d25}. The output state of the standard MZI\ with the internal phonton
losses is given by $\left \vert \psi \right \rangle _{out}=\hat{B}_{2}\hat{U}%
_{\phi }\hat{B}_{Lw}\hat{B}_{1}\left \vert \psi \right \rangle _{in}\otimes
\left \vert 0\right \rangle _{a_{v}}\otimes \left \vert 0\right \rangle
_{b_{v}}$.
\begin{figure}[tbp]
\label{Fig1} {\centering \includegraphics[width=0.95\columnwidth]{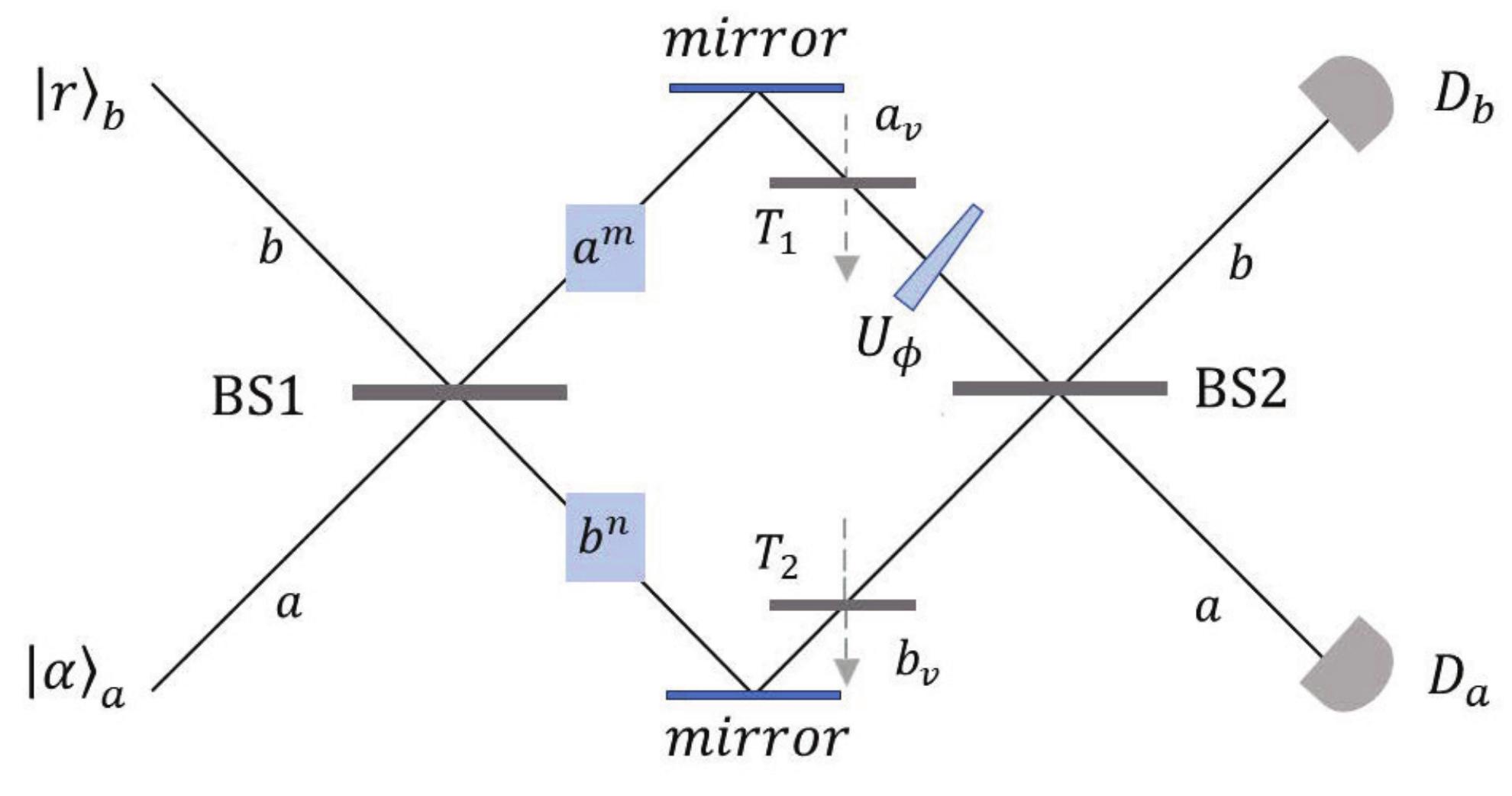}}
\caption{Schematic diagram of a MZI with multi-PSSs. The two input ports are
a coherent state $\left \vert \protect \alpha \right \rangle _{a}$ and a
squeezed vacuum state $\left \vert r\right \rangle _{b}$, respectively. $BS$
is the beamsplitter, $U_{\protect \phi }$ is the phase shifter, and $D_{a}$ ($%
D_{b}$) is the specific detector. $a^{m}$ represents the operation of
subtracting $m$ photons on mode $a$. $b^{n}$ represents the operation of
subtracting $n$ photons on mode $b$.}
\label{1}
\end{figure}

To enhance phase sensitivity, we introduce the photon subtraction operations
inside the MZI after the first BS, as illustrated in Fig.\ 1. The photon
subtraction has experimentally proved to be feasible. In Ref. \cite{d26},
they realized a single-photon subtraction experimentlly by using high
transsivity BS. Theoretically, the multi-photon subtraction can be realized
via detecting $m$ photons after splitting the field by a BS of high
transmittivity \cite{e11,d21}. In our scheme, $m$ and $n$ photons are
subtracted from mode $a$ and mode $b$, respectively. This process can be
described as $\hat{a}^{m}\hat{b}^{n}$. And we utilize a coherent state $%
\left \vert \alpha \right \rangle _{a}=\hat{D}(\alpha )\left \vert
0\right
\rangle _{a}$ and a squeezed vacuum state $\left \vert r\right
\rangle _{b}=\hat{S}_{b}(r)\left \vert 0\right \rangle _{b}$\ as input
states, where $\hat{D}(\alpha )=\exp (\alpha \hat{a}^{\dagger }-\alpha
^{\ast }\hat{a})$ is the translation operator with the translation parameter
$\alpha $ ($\alpha =\left \vert \alpha \right \vert e^{i\theta _{\alpha }}$)
and $\hat{S}_{b}(r)=\exp [r(\hat{b}^{2}-\hat{b}^{\dagger 2})/2]$ is the
single-mode squeezing operator with the squeezing parameter $r$. Thus, in
the extended space with photon losses, the output state of the
interferometer can be expressed as a pure state form as

\begin{equation}
\left \vert \Psi \right \rangle _{out}=A\hat{B}_{2}\hat{U}_{\phi }\hat{B}%
_{Lw}\hat{a}^{m}\hat{b}^{n}\hat{B}_{1}\left \vert \alpha \right \rangle
_{a}\left \vert r\right \rangle _{b}\left \vert 0\right \rangle
_{a_{v}}\left \vert 0\right \rangle _{b_{v}},  \label{b0}
\end{equation}%
where $A$ is the normalization coefficient.

It is convenient to denote $\left \langle \hat{a}^{\dagger p_{1}}\hat{a}%
^{p_{2}}\hat{b}^{\dagger q_{1}}\hat{b}^{q_{2}}\right \rangle $ as universal
formula. According to Eq. (\ref{b0}), one can obtain the universal formula,
i.e.,%
\begin{equation}
\left \langle \hat{a}^{\dagger p_{1}}\hat{a}^{p_{2}}\hat{b}^{\dagger q_{1}}%
\hat{b}^{q_{2}}\right \rangle =A^{2}D_{m,n,p_{1},p_{2},q_{1},q_{2}}e^{M},
\label{b1}
\end{equation}%
where%
\begin{align}
& D_{m,n,p_{1},p_{2},q_{1},q_{2}}  \notag \\
& =\frac{\partial ^{p_{1}+p_{2}+q_{1}+q_{2}}}{\partial x_{1}^{p_{1}}\partial
y_{1}^{q_{1}}\partial y_{2}^{q_{2}}\partial x_{2}^{p_{2}}}\frac{\partial
^{2m+2n}}{\partial s_{1}^{m}\partial t_{1}^{n}\partial s_{2}^{m}\partial
t_{2}^{n}}\left \{ \cdot \right \}  \notag \\
& |_{x_{1}=x_{2}=y_{1}=y_{2}=s_{1}=t_{1}=s_{2}=t_{2}=0},  \label{b2}
\end{align}%
and%
\begin{eqnarray}
M &=&M_{1}\alpha ^{\ast }+M_{4}\alpha +M_{2}M_{3}\sinh ^{2}r  \notag \\
&&-\frac{1}{2}\cosh r\sinh r\left( M_{2}^{2}+M_{3}^{2}\right) ,  \label{01}
\end{eqnarray}%
with%
\begin{eqnarray}
M_{1} &=&\frac{s_{1}+it_{1}}{\sqrt{2}}+\sqrt{T}(\frac{1+e^{-i\phi }}{2}%
x_{1}+i\frac{1-e^{-i\phi }}{2}y_{1}),  \label{02} \\
M_{2} &=&\frac{t_{1}+is_{1}}{\sqrt{2}}+\sqrt{T}(\frac{1+e^{-i\phi }}{2}%
y_{1}-i\frac{1-e^{-i\phi }}{2}x_{1}),  \label{03} \\
M_{3} &=&\frac{t_{2}-is_{2}}{\sqrt{2}}+\sqrt{T}(\frac{1+e^{i\phi }}{2}y_{2}+i%
\frac{1-e^{i\phi }}{2}x_{2}),  \label{04} \\
M_{4} &=&\frac{s_{2}-it_{2}}{\sqrt{2}}+\sqrt{T}(\frac{1+e^{i\phi }}{2}x_{2}-i%
\frac{1-e^{i\phi }}{2}y_{2}).  \label{05}
\end{eqnarray}%
Thus, the normalization coefficient for the multi-PSSs is given by%
\begin{equation}
A=\frac{1}{\sqrt{D_{m,n,0,0,0,0}e^{M}}},  \label{06}
\end{equation}%
where $D_{m,n,0,0,0,0}=\frac{\partial ^{2m+2n}}{\partial s_{1}^{m}\partial
t_{1}^{n}\partial s_{2}^{m}\partial t_{2}^{n}}\left \{ \cdot \right \}
|_{s_{1}=t_{1}=s_{2}=t_{2}=0}$.

Here we will briefly introduce the photon subtraction operations of our
scheme. In this paper, these operations performed within MZI can be divided
into three operation schemes as follows: (i) Scheme A, set $n=0$,
subtracting $m$ photons from mode $a$, i.e., $\hat{a}^{m}$; (ii) Scheme B,
set $m=0$, subtracting $n$ photons from mode $b$, i.e., $\hat{b}^{n}$; (iii)
Scheme C is the successive implementation of scheme A and scheme B, i.e., $%
\hat{a}^{m}\hat{b}^{n}$.

\section{Phase sensitivity}

The phase sensitivity is a key parameter for measuring the unknown phase
accuracy of an optical interferometer, which is closely related to the
specific detection method \cite{d28,d29}. The smaller the value of phase
sensitivity, the higher the corresponding phase accuracy. Measurement of the
same interferometric output field by different detection methods will result
in different phase sensitivities. Common detection methods include homodyne
detection \cite{d3,d5,d7}, intensity detection \cite{d8,d9} and parity
detection \cite{d11,d12}. However, many studies have shown that the parity
detection is harder to implement experimentally and more susceptible to
losses. Therefore, we only compare the intensity detection and homodyne
detection methods. Next, we will discuss the effect of multi-PSSs on the
phase sensitivity based on intensity detection and homodyne detection
methods.

According to the error propagation equation \cite{c10}, the phase
sensitivity can be expressed as:%
\begin{equation}
\Delta \phi =\frac{\sqrt{\left \langle \Delta ^{2}\hat{O}_{k}\right \rangle }%
}{\left \vert \partial _{\phi }\left \langle \hat{O}_{k}\right \rangle
\right \vert },  \label{s1}
\end{equation}%
where $\hat{O}_{k}$ is the operator corresponding the selected measurement ($%
\hat{O}_{1}=c_{1}\hat{N}_{a}+d_{1}\hat{N}_{b}$, $\hat{O}_{2}=c_{2}\hat{X}%
_{a}+d_{2}\hat{X}_{b}$), $\Delta ^{2}\hat{O}_{k}=\left \langle \hat{O}%
_{k}^{2}\right \rangle -\left \langle \hat{O}_{k}\right \rangle ^{2}$, and $%
\partial _{\phi }\left \langle \hat{O}_{k}\right \rangle =\partial \left
\langle \hat{O}_{k}\right \rangle /\partial \phi $. According to Eqs. (\ref%
{b1}) and (\ref{s1}), one can obtain the phase sensitivity for our scheme in
principle.

\subsection{The optical intensity detection}

First, we briefly introduce the operator $\hat{O}_{1}$ corresponding to the
intensity detection methods, i.e.,%
\begin{equation}
\hat{O}_{1}=c_{1}\hat{N}_{a}+d_{1}\hat{N}_{b},  \label{s2}
\end{equation}%
where $\hat{N}_{a}=\hat{a}^{\dagger }\hat{a}$ and $\hat{N}_{b}=\hat{b}%
^{\dagger }\hat{b}$ are the particle number operators of the output port $a$
and the output port $b$, respectively. $c_{1}$ and $d_{1}$ are adjustable
coefficients.

Intensity detection is a measurement of the photocurrent. We consider three
types of intensity detection, including single-intensity detection
(intensity detection on mode $a$ or $b$, i.e., $N_{a}$ or $N_{b}$) and
intensity difference detection ($N_{-}$). The choice of the above detection
methods depends on the values of $c_{1}$ and $d_{1}$ as follows: (i) when $%
c_{1}=1$ and $d_{1}=0$, corresponding to $N_{a}$ (single-intensity detection
on mode $a$), the phase sensitivity is $\Delta \phi _{n_{a}}$; (ii) when $%
c_{1}=0$ and $d_{1}=1$, corresponding to $N_{b}$ (single-intensity detection
on mode $b$), the phase sensitivity is $\Delta \phi _{n_{b}}$; (iii) when $%
c_{1}=1$ and $d_{1}=$ $-1$, corresponding to $N_{-}$ (intensity difference
detection), the phase sensitivity is $\Delta \phi _{n_{-}}$. According to
Eqs. (\ref{b0}), (\ref{s1}) and (\ref{s2}), we can obtain the phase
sensitivity for multi-PSSs. The calculation process is provided in Appendix
A.

Next, to analyze which operation schemes perform best and to find the
optimal intensity detection method, we plot the phase sensitivity as a
function of $\phi $ based on these intensity detection methods. For
simplicity, we examine the effect of different multi-PSSs under the
condition that the number of photons subtracted is fixed at 2, i.e., scheme
A ($m=2,n=0$) indicates two photons subtracted in mode $a$, scheme B ($%
m=0,n=2$) denotes two photons subtracted in mode $b$, and scheme C ($m=1,n=1$%
) signifies one photon being subtracted from each of two modes.
Additionally, the standard scheme ($m=0,n=0$) is a standard MZI without
multi-PSSs.
\begin{figure}[tbp]
\label{Fig2} {\centering \includegraphics[width=0.95\columnwidth]{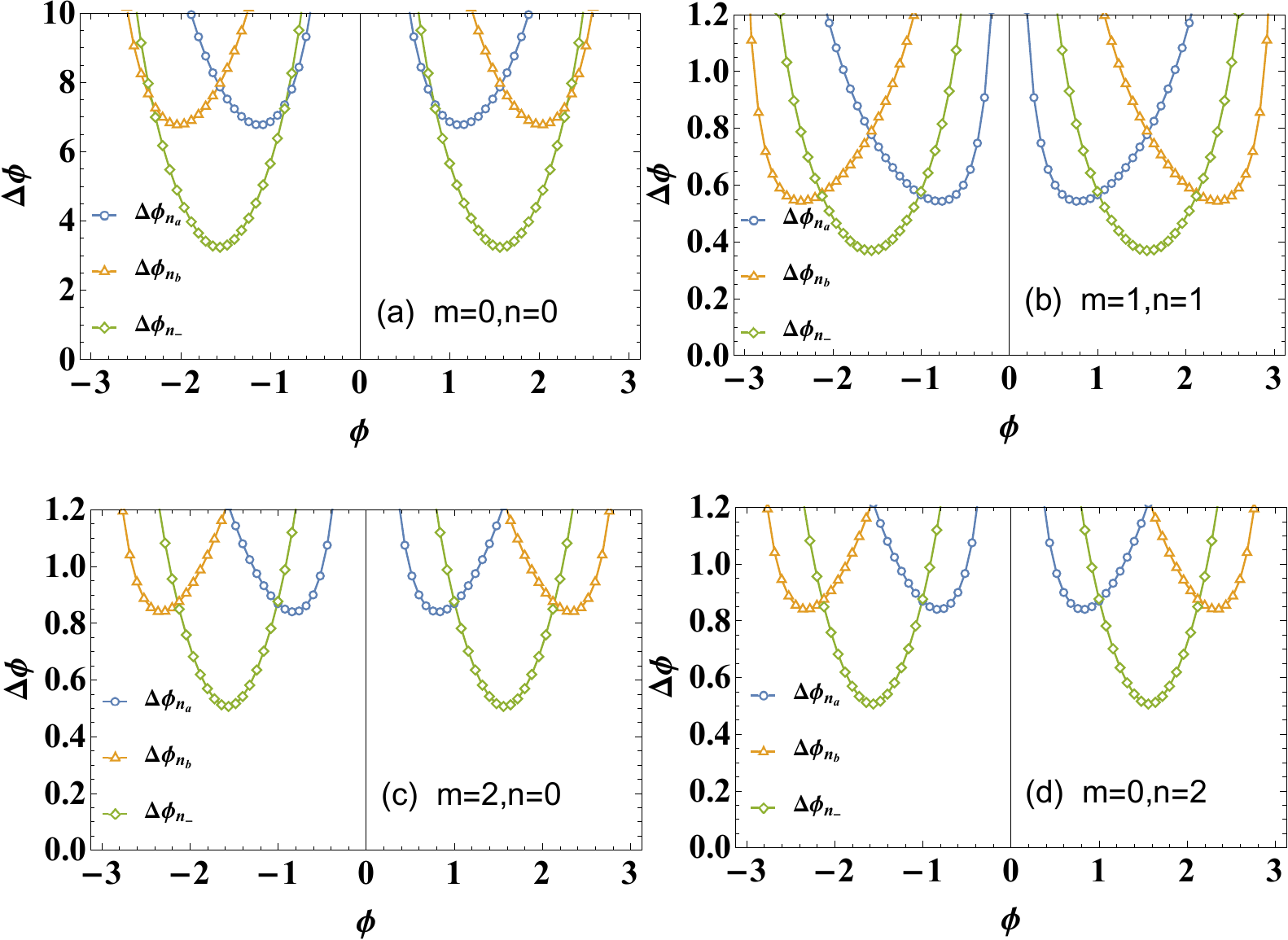}}
\caption{The phase sensitivity of multi-PSSs based on intensity detection as
a function of $\protect \phi $ with $\protect \alpha $ $=1$, $r=1$ and $T=1$.
Including single mode multi-PSS: scheme A ($m$=2,$n$=0) and scheme B ($m$=0,$%
n$=2) and symmetrical two-mode multi-PSS: scheme C ($m$=$n$=1).}
\label{2}
\end{figure}

In Fig. 2, it can be clearly seen that among the intensity detection
methods, (i) the intensity difference detection $N_{-}$ is the best among
these, superior to single-intensity detection ($N_{a}$ and $N_{b}$).
Additionally, there is almost no difference between $N_{a}$ and $N_{b}$.
(ii) These operation schemes of photon subtraction have obvious improvement
in phase sensitivity by using intensity detection. (iii) Scheme C ($m=1,n=1$%
) has the best effect on phase sensitivity improvement (refer to Fig. 2(b)).
The enhancement achieved by scheme A ($m=2,n=0$) is identical to that of
scheme B ($m=0,n=2$).

That is to say, the optimal intensity detection method is $N_{-}$ and the
best scheme based on intensity detection methods is scheme C, followed by
schemes A and B.

\subsubsection{Phase sensitivity based on intensity difference detection}

Subsequently, we examine the phase sensitivity with scheme A based on
intensity difference detection $N_{-}$, focusing on the effects of several
parameters such as the phase, the number of photons subtracted ($m$), the
coherent amplitude $\alpha $, and the squeezing parameter $r$. In order to
facilitate analysis, we compare the performance of the three multi-PSSs in
phase estimation by subtracting the same number of photons from mode $a$.

\paragraph{Ideal case}

First, we consider the ideal case, corresponding to $T_{k}=1$. In Fig. 3, we
respectively plot phase sensitivity based on the intensity difference
detection as a function of the phase, coherent amplitude $\alpha $, and the
squeezing parameter $r$. As shown in Fig. 3, we can clearly observe that,
based on the intensity difference detection $N_{-}$ with scheme A: (i) the
phase sensitivity $\Delta \phi _{n_{-}}$ can be improved with increasing $m$%
, and it reaches its optimal value at approximately $\phi =1.6$. (ii) $%
\Delta \phi _{n_{-}}$ demonstrates a trend of initially increasing followed
by a decrease as $\alpha $ increases. Notably, in the smaller, rather than
larger parameter range, we can see an enhancement of the phase sensitivity.
(iii) $\Delta \phi _{n_{-}}$ shows a trend of initially increasing followed
by a decrease as $r$ increases; however, significant enhancements occur only
within the larger parameter range of $r$.
\begin{figure}[tbp]
\label{Fig3} {\centering \includegraphics[width=0.95\columnwidth]{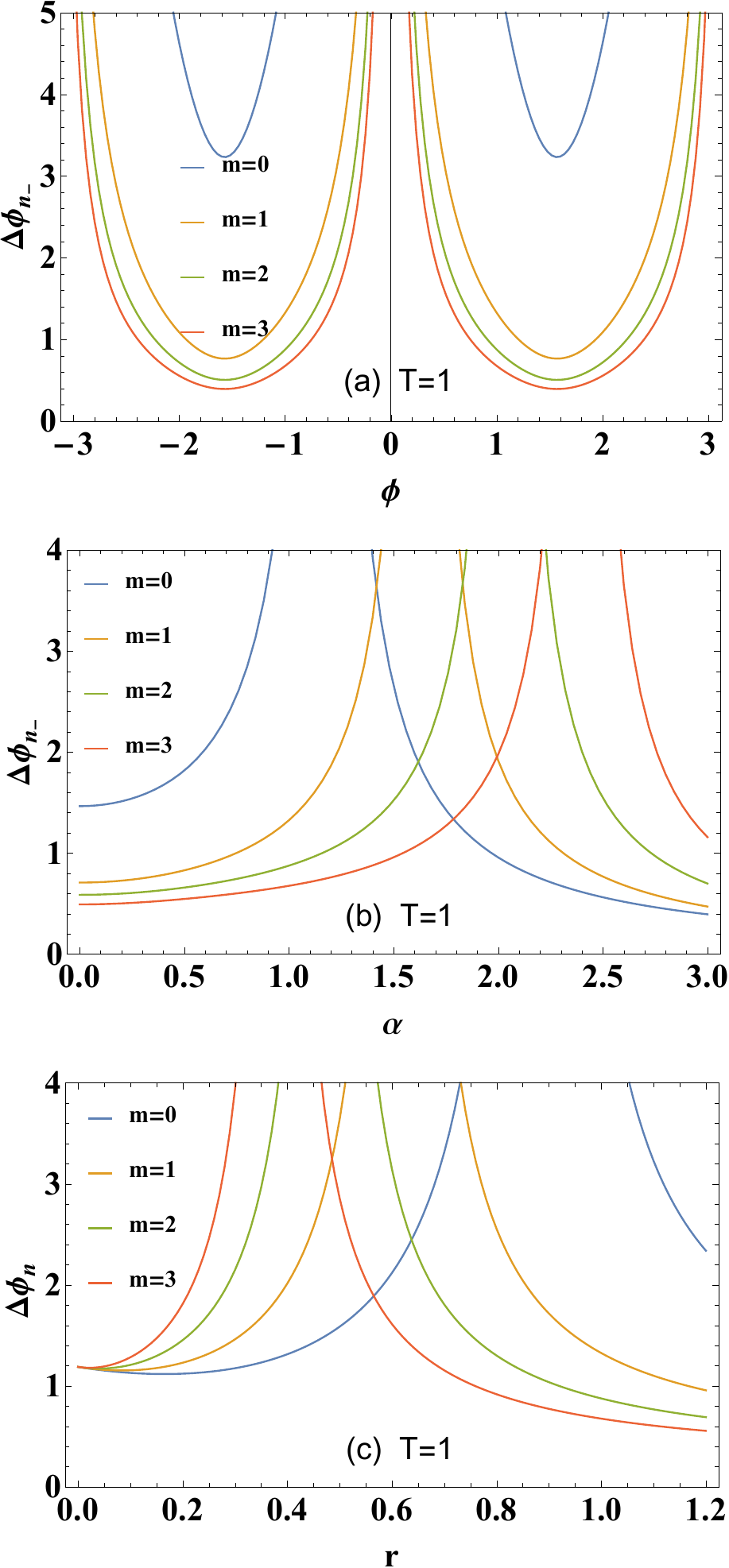}}
\caption{The phase sensitivity of scheme A based on $N_{-}$ as a function of
(a) phase $\protect \phi $ with $\protect \alpha =1$, and $r=1$; (a) the
coherent amplitude $\protect \alpha $, with $r=1$ and $\protect \phi =1$; (b)
the squeezing parameter $r$, with $\protect \alpha =1$ and $\protect \phi =1$.}
\end{figure}

\paragraph{Photon losses case}

In practical situations, quantum measurement needs to take into account the
influence of the environment, especially that of the photon losses inside
the interferometer. Fig. 4 shows the results with photon losses ($0<T<1$).
We plot the phase sensitivity as a function of the transmittance $T$ for
fixed $r$, $\alpha $, $\phi $, and photon-subtracted numbers. As shown in
Fig. 4: (i) With other parameters fixed, the phase sensitivity decreases as
the transmittance $T$ decreases, as expected. (ii) It can be seen that as
the losses increase, the curve of the standard scheme drops more steeply and
shows higher sensitivity to changes in $T$, while the curve of scheme A
changes relatively gently. This indicates that our scheme has stronger
anti-loss ability and robustness than the standard scheme with internal
photon losses.
\begin{figure}[tbp]
\label{Fig4} {\centering \includegraphics[width=0.95\columnwidth]{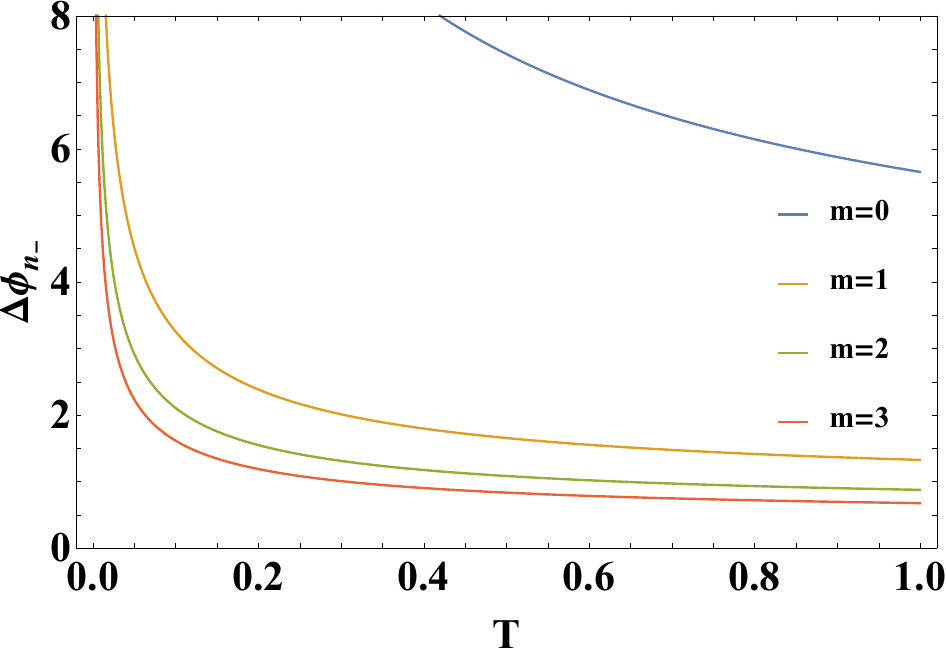}}
\caption{The phase sensitivity of scheme A as a function of transmittance $T$%
, with $\protect \alpha =1,r=1$ and $\protect \phi =1$.}
\label{4}
\end{figure}

\paragraph{Comparison with SQL and HL}

Furthermore, we compare the phase sensitivity with the SQL and the HL. The
SQL and HL are respectively defined as $\Delta \phi _{SQL}$ $=1/\sqrt{N}$
and $\Delta \phi _{HL}$ $=1/N$, where $N$ is the total mean photon number
inside the MZI before the second BS \cite{d14,d15,d16}. For multi-PSSs, $N$
can be calculated as%
\begin{eqnarray}
N &=&A^{2}\left \langle \psi \right \vert _{in}\hat{B}_{1}^{\dag }\hat{a}%
^{\dag m}\hat{b}^{\dag n}(\hat{a}^{\dag }\hat{a}+\hat{b}^{\dag }\hat{b})\hat{%
a}^{m}\hat{b}^{n}\hat{B}_{1}\left \vert \psi \right \rangle _{in}  \notag \\
&=&A^{2}(\left \langle \psi \right \vert _{in}\hat{B}_{1}^{\dag }\hat{a}%
^{\dag m+1}\hat{b}^{\dag n}\hat{a}^{m+1}\hat{b}^{n}\hat{B}_{1}\left \vert
\psi \right \rangle _{in}  \notag \\
&&+\left \langle \psi \right \vert _{in}\hat{B}_{1}^{\dag }\hat{a}^{\dag m}%
\hat{b}^{\dag n+1}\hat{a}^{m}\hat{b}^{n+1}\hat{B}_{1}\left \vert \psi \right
\rangle _{in}).  \label{ss2}
\end{eqnarray}

For convenience, we calculate the general formula $\left. _{in}\left \langle
\psi \right \vert B_{1}^{\dagger }\hat{a}^{\dagger m_{1}}\hat{b}^{\dagger
n_{1}}\hat{b}^{n_{2}}\hat{a}^{m_{2}}B_{1}\left \vert \psi \right \rangle
_{in}\right. $, which has the mathematical analytic form, i.e.,%
\begin{equation}
\left. _{in}\left \langle \psi \right \vert B_{1}^{\dagger }\hat{a}^{\dagger
m_{1}}\hat{b}^{\dagger n_{1}}\hat{b}^{n_{2}}\hat{a}^{m_{2}}B_{1}\left \vert
\psi \right \rangle _{in}\right. =D_{m_{1},m_{2},n_{1},n_{2}}e^{Q},
\label{s7}
\end{equation}%
where%
\begin{equation}
D_{m_{1},m_{2},n_{1},n_{2}}=\frac{\partial ^{m_{1}+n_{1}+m_{2}+n_{2}}}{%
\partial s_{1}^{m_{1}}\partial t_{1}^{n_{1}}\partial t_{2}^{n_{2}}\partial
s_{2}^{m_{2}}}\left \{ \cdot \right \} |_{s_{1}=t_{1}=s_{2}=t_{2}=0},
\label{s8}
\end{equation}%
and
\begin{eqnarray}
Q &=&\frac{1}{\sqrt{2}}\left( s_{1}+it_{1}\right) \alpha ^{\ast }+\frac{1}{%
\sqrt{2}}\left( s_{2}-it_{2}\right) \alpha  \notag \\
&&+\frac{1}{2}\left( t_{1}+is_{1}\right) \left( t_{2}-is_{2}\right) \sinh
^{2}r  \notag \\
&&-\frac{1}{4}\cosh r\sinh r[\left( t_{1}+is_{1}\right) ^{2}+\left(
t_{2}-is_{2}\right) ^{2}].  \label{s9}
\end{eqnarray}%
Here, $m_{1},m_{2},n_{1},n_{2}$ are integers, $s_{1},s_{2},t_{1},t_{2}$ are
differential variables,\ and\ after\ the\ differentiation,\ all\
differential variables are taken to be zero.

According to Eq. (\ref{ss2}) and (\ref{s7}), the total mean photon number $N$
for scheme A is given by%
\begin{equation}
N=4A^{2}(D_{m+1,m+1,n,n}+D_{m,m,n+1,n+1})e^{Q}.  \label{ss3}
\end{equation}%
\begin{figure}[tbp]
\label{Fig5} {\centering \includegraphics[width=0.95\columnwidth]{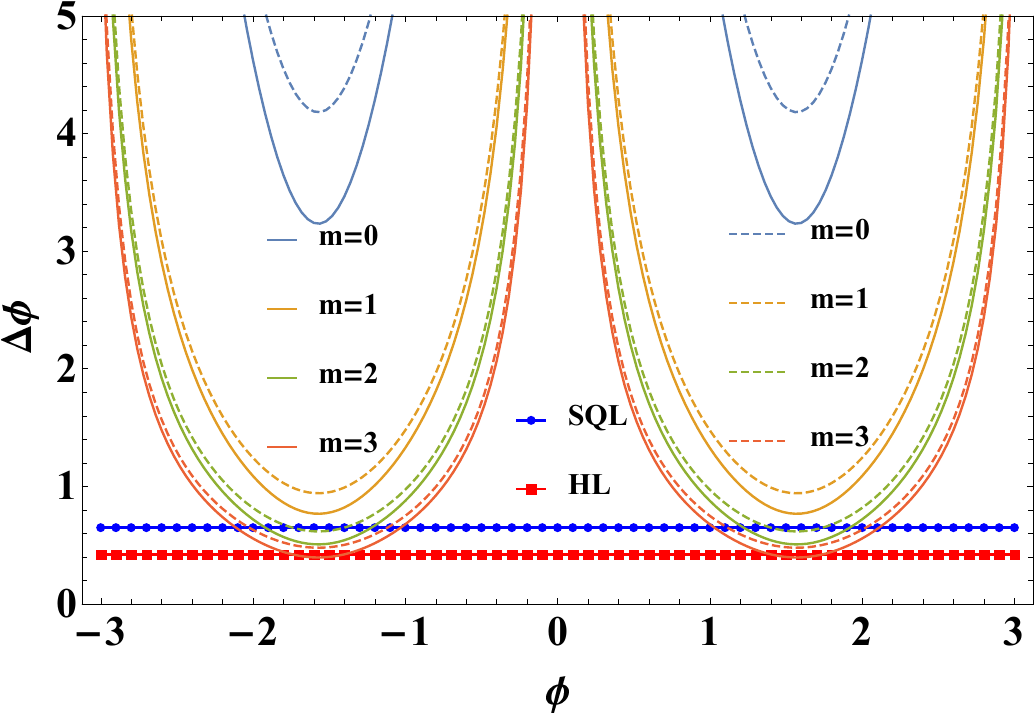}}
\caption{Comparison of the phase sensitivity of scheme A based on $N_{-}$
with SQL and HL. The solid blue circles represent SQL, and the solid red
squares represent HL. The blue solid line corresponds to the standard
scheme, the yellow, the green and the red solid line correspond to the
simultaneous subtraction of one photon, two photons and three photons from
mode $a$, respectively.}
\label{5}
\end{figure}

With fixed $\alpha $ and $r$, we compare the phase sensitivity with the SQL
and HL, as shown in Fig. 5. It is found that: (i) In ideal cases, the
standard scheme ($m=0$) and single-photon subtraction operation ($m=1$)
cannot break the SQL, while multi-photon subtraction operations for scheme A
($m=2,3$) can break the SQL within a wide range. Note that when $m=3$, it
can even break the HL (Fig. 6(a)). (ii) Under relatively large internal
photon losses ($T=0.7$), scheme A ($m=2,3$) can even break through the SQL,
as shown in Fig. 6(b). This implies that our scheme exhibits robustness
against internal photon losses.

\subsection{The optical homodyne detection}

Here, we introduce the operator $\hat{O}_{2}$ corresponding to the homodyne
detection methods, which is expressed as:%
\begin{equation}
\hat{O}_{2}=c_{2}\hat{X}_{a}+d_{2}\hat{X}_{b},  \label{s4}
\end{equation}%
where $\hat{X}_{a}=(\hat{a}+\hat{a}^{\dagger })/\sqrt{2}$ and $\hat{X}_{b}=(%
\hat{b}+\hat{b}^{\dagger })/\sqrt{2}$ are the orthogonal component operators
of the output ports $a$ and $b$, respectively. $c_{2}$ and $d_{2}$ are
adjustable coefficients.

Homodyne detection methods include: (i) When $c_{2}=1$ and $d_{2}=0$, $a$
mode is detected, i.e., $X_{a}$, whose phase sensitivity is $\Delta \phi
_{X_{a}}$; (ii) when $c_{2}=0$ and $d_{2}=1$, $b$ mode is detected, i.e., $%
X_{b}$, whose phase sensitivity is $\Delta \phi _{X_{b}}$. According to Eqs.
(\ref{b0}), (\ref{s1}) and (\ref{s4}), we can obtain the expressions for the
phase sensitivities based on homodyne detection methods. The details are
given in Appendix A.

Next, we will explore the determination of the optimal homodyne detection
method, as well as identify the best multi-PSS in order to enhance the phase
sensitivity, maximally. Thus, we plot the phase sensitivity $\Delta \phi $
as a function of $\phi $ based on the homodyne detection in Fig.6, including
the standard scheme (Fig. 6(a)), scheme A and B (Fig. 6(c) and (d)) and
scheme C (Fig. 6(b)).
\begin{figure}[tbp]
\label{Fig6} {\centering \includegraphics[width=0.95\columnwidth]{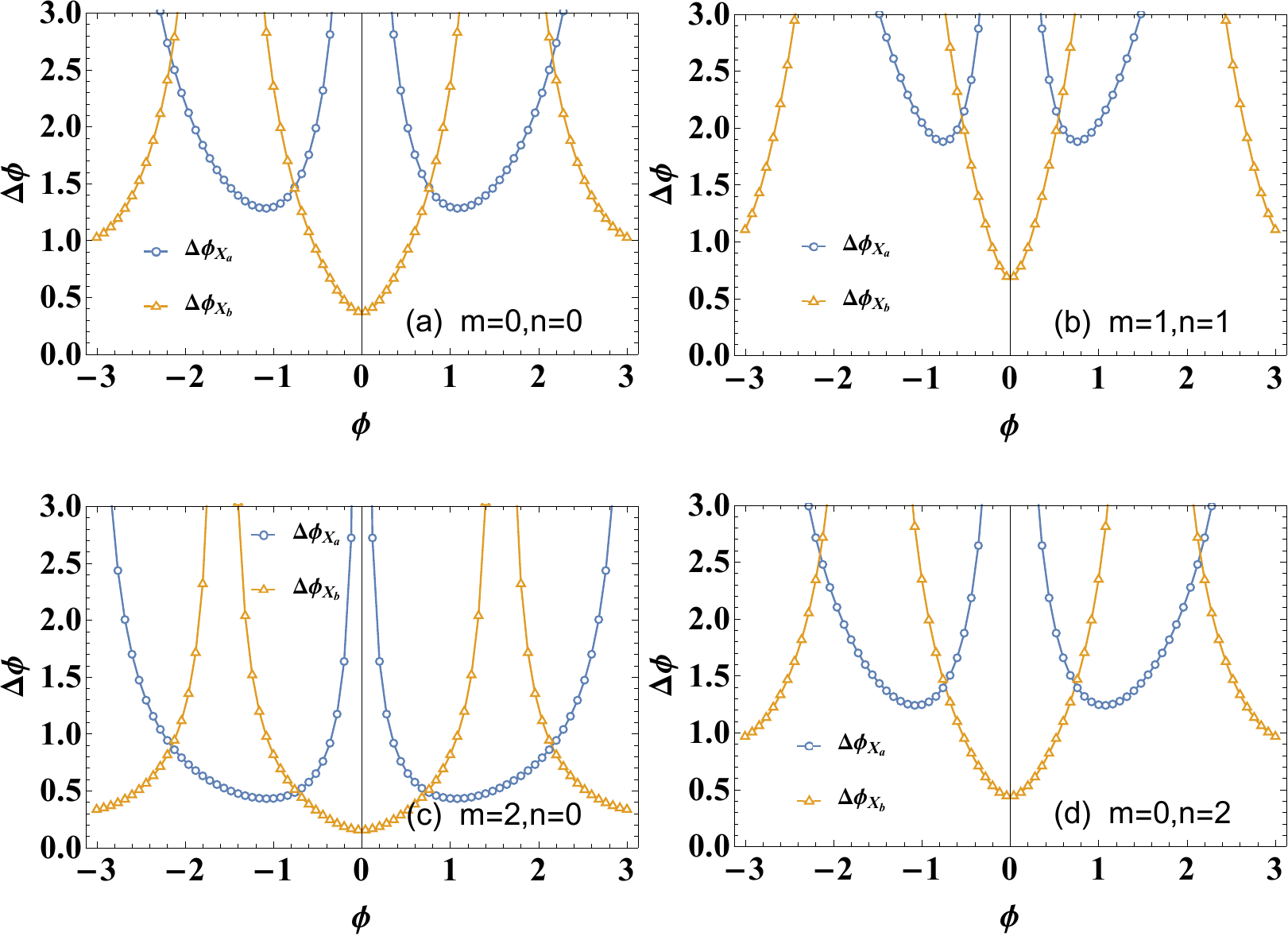}}
\caption{The phase sensitivity of the multi-PSSs based on the homodyne
detection as a function of $\protect \phi $ with $\protect \alpha =1$ and $r=1$
and $T=1$. Including single mode multi-PSS: scheme A ($m$=2,$n$=0) and
scheme B ($m$=0,$n$=2) and symmetrical two-mode multi-PSS: scheme C ($m$=$n$%
=1).}
\label{6}
\end{figure}

From Fig. 6, it is evident that: (i) the phase sensitivity based on $X_{b}$
performs better than that based on $X_{a}$. (ii) With fixed parameters $%
\alpha =1,r=1$, different multi-PSSs within MZI does not always enhance the
phase sensitivity. Specifically, scheme A ($m=2,n=0$) significantly improves
phase sensitivity. However, the influence of scheme B ($m=0,n=2$) on phase
sensitivity is not remarkable and scheme C does even degrade the phase
sensitivity. This indicates the optimal homodyne detection scheme is $X_{b}$
with scheme A as its best scheme.

\subsubsection{Phase sensitivity based on homodyne detection $X_{b}$}

Now, we examine the phase sensitivity with scheme A based on homodyne
detection $X_{b}$ focusing on the influence of its associated parameters.

\paragraph{Ideal case}

We analyze the effects of the photons subtracted number $m$, coherent state
amplitude, and squeezing parameter on the phase sensitivity. In Fig. 7, we
plot the phase sensitivity $\Delta \phi $ based on $X_{b}$ with scheme A as
a function of $\phi $, $\alpha $ and $r$, respectively. From Fig. 7, it is
evident that, based on $X_{b}$, (i) the phase sensitivity can be improved
with increasing $m$. Additionally, the phase sensitivity reaches its optimum
value at $\phi =0$; (ii) further enhancement in phase sensitivity occurs
with an increase in $\alpha $ and $r$. So $m$, $\alpha $ and $r$ all
contribute positively to enhancing phase sensitivity in this scheme.
\begin{figure}[tbp]
\label{Fig7} {\centering \includegraphics[width=0.95\columnwidth]{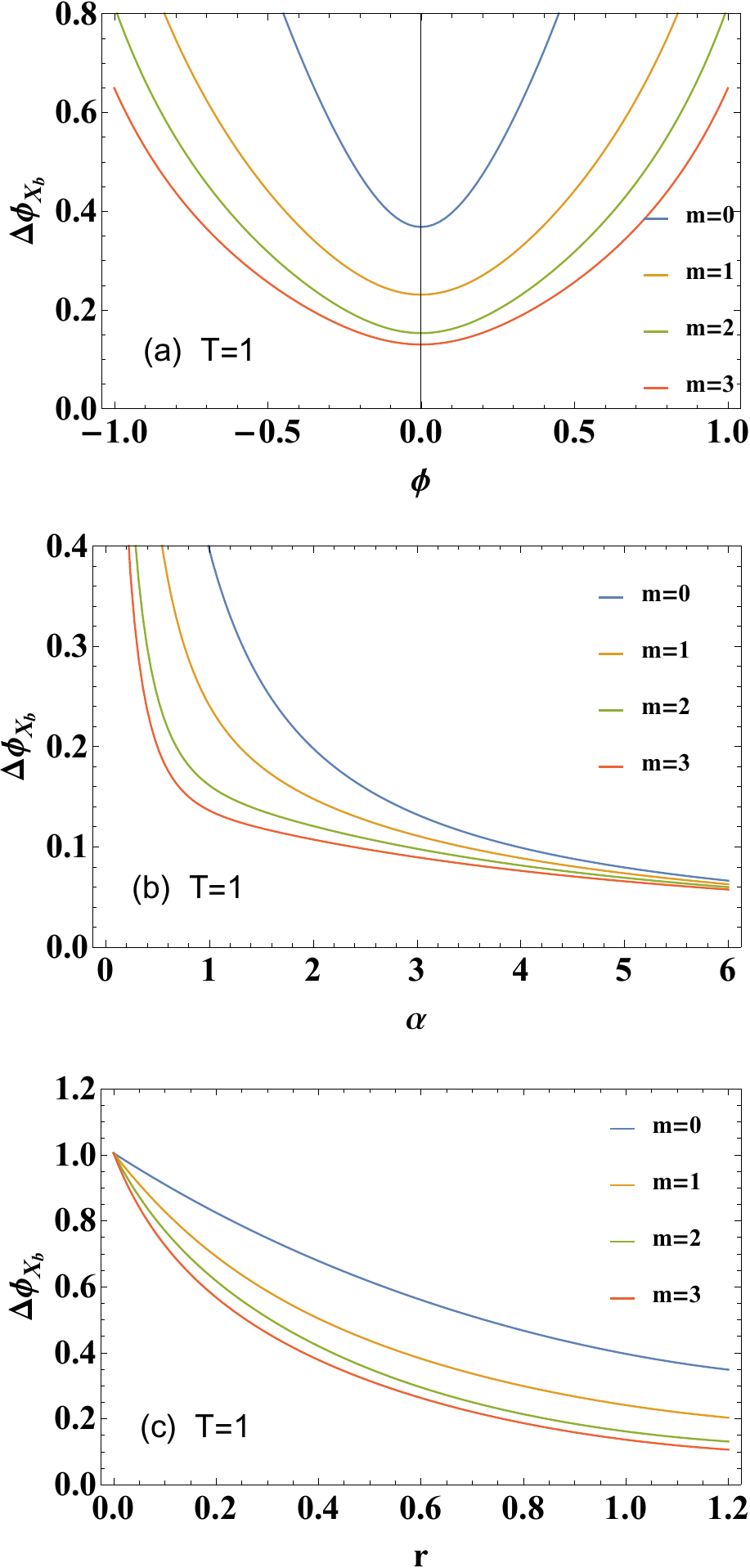}}
\caption{The phase sensitivity of scheme A based the homodyne detection $%
X_{b}$ as a function of (a) phase $\protect \phi $ with $\protect \alpha =1$,
and $r=1$, (b) the coherent amplitude $\protect \alpha $, with $r=1$, and $%
\protect \phi =0.1$, (c) the squeezing parameter $r$, with $\protect \alpha =1$
and $\protect \phi =0.1$.}
\label{7}
\end{figure}

\paragraph{Photon losses case}

In order to demonstrate how the phase sensitivity of scheme A based on $X_{b}
$ behaves in the lossy case, we plot the phase sensitivity as a function of
parameters such as transmittance, coherence amplitude, and squeezing
parameter for the lossy case. As shown in Fig. 8, the obtained results are
similar to those with the intensity difference detection method, and the
phase sensitivity decreases with the decrease of transmittance $T$. The
curve of the standard scheme varies significantly more steeply with the
parameter $T$, compared to the curve in scheme A. This indicates that the
standard scheme has weaker robustness, while scheme A shows stronger
robustness.
\begin{figure}[tbp]
\label{Fig8} {\centering \includegraphics[width=0.95\columnwidth]{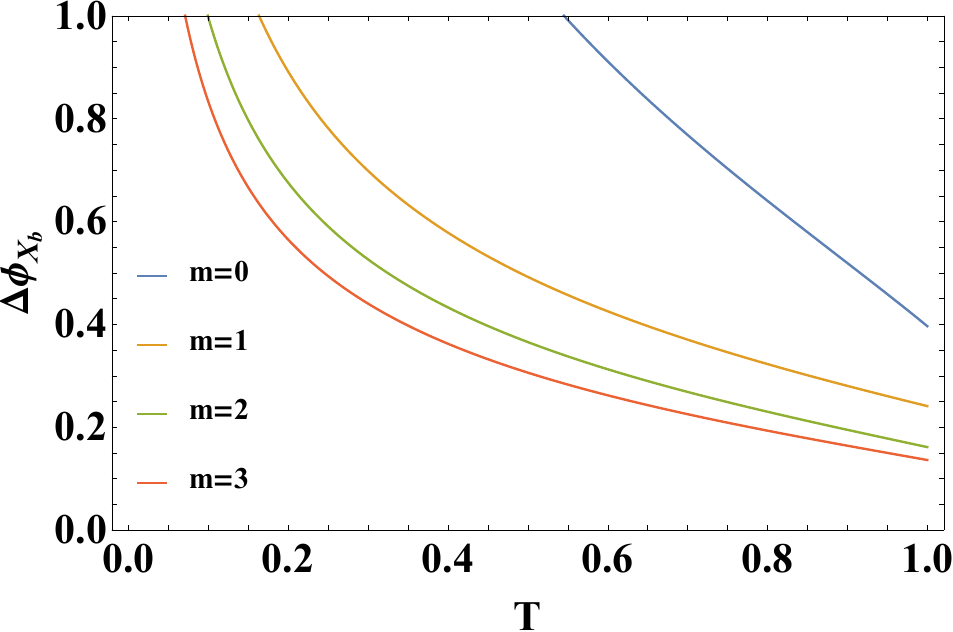}}
\caption{The phase sensitivity of scheme A as a function of transmittance $T$%
, with $\protect \alpha =1$ and $r=1$.}
\label{8}
\end{figure}

\paragraph{Comparison with SQL and HL}

As shown in Fig. 9, the phase sensitivity in ideal case has already broken
through the HL, indicating that the detection method $X_{b}$ has a good
advantage under scheme A. At the same time, it can be seen that scheme A
breaks through the HL in a larger width range, which has a significant
improvement on the phase sensitivity. At the high loss of $T=0.7$, our
scheme (based on homodyne detection) can still break through the HL and has
a wider range as $m$ increases, while the standard scheme does not even
manage to break through the SQL. This shows that our scheme has good
robustness.
\begin{figure}[tbp]
\label{Fig9} {\centering \includegraphics[width=0.95\columnwidth]{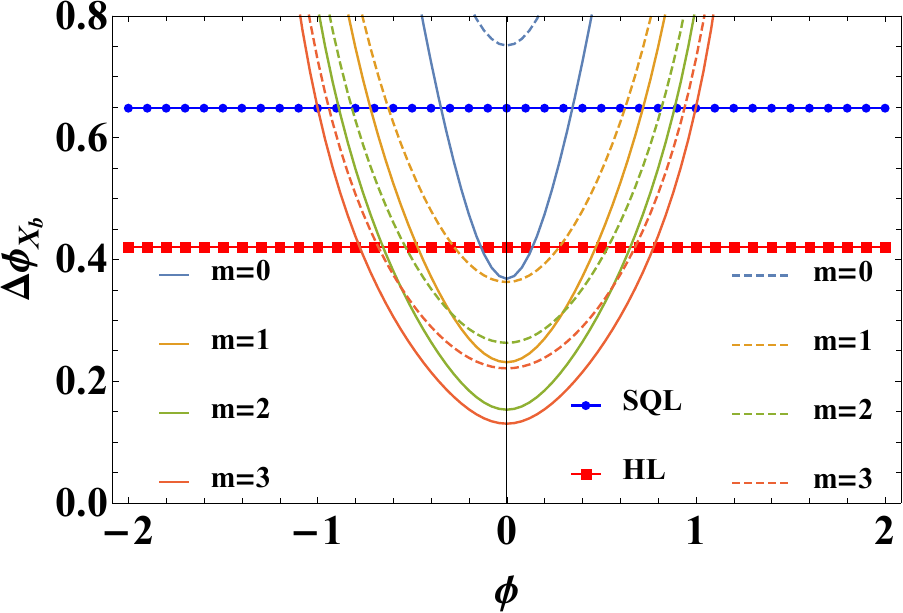}}
\caption{Comparison of the phase sensitivity of scheme A based on $X_{b}$
with SQL and HL. The solid blue circles is SQL, and the solid red squares is
HL. The blue solid line corresponds to the standard MZI, the yellow, the
green and the red solid lines correspond to the simultaneous deduction of
one photon, two photons and three photons from mode $a$, respectively.}
\label{9}
\end{figure}

In a word, scheme A improves the phase sensitivity significantly, and both
of them break through the HL, which indicates that this detection method is
better than the intensity difference detection scheme, and it is the optimal
detection method in the presence of losses.

\section{The QFI}

The phase sensitivity is determined based on error propagation formula and
must be influenced by the specific measurement preferences. Additionally, a
theoretical framework for optimal phase sensitivities is required, which can
be derived from the Fisher information according to the QCRB theory. The QFI
constitutes a highly efficacious approach to identifying the optimal
solution for parameter estimation, which could represent the theoretical
maximum information of unknown phase shift.

\subsection{Ideal case}

When the system is in the lossless scenario, for pure input states, the QFI
is \cite{f0}%
\begin{equation}
F_{Q}=4\left[ \left \langle \psi _{\phi }^{\prime }\right \vert \left \vert
\psi _{\phi }^{\prime }\right \rangle -\left \vert \left \langle \psi _{\phi
}^{\prime }\right \vert \left \vert \psi _{\phi }\right \rangle \right \vert
^{2}\right] ,  \label{s5}
\end{equation}%
where $\left \vert \psi _{\phi }\right \rangle $ is the quantum state after
the phase shift and before the second BS, and $\left \vert \psi _{\phi
}^{\prime }\right \rangle =\partial _{\phi }\left \vert \psi _{\phi
}\right
\rangle =\partial \left \vert \psi _{\phi }\right \rangle /\partial
\phi $. Then the QFI can be rewritten as
\begin{equation}
F=4\left \langle \Delta ^{2}\hat{N}_{a}\right \rangle ,  \label{s6}
\end{equation}%
where $\left \langle \Delta ^{2}\hat{N}_{a}\right \rangle =\left \langle
\psi _{\phi }\right \vert (\hat{a}^{\dagger }\hat{a})^{2}|\psi _{\phi
}\rangle -(\left \langle \psi _{\phi }\right \vert \hat{a}^{\dagger }\hat{a}%
|\psi _{\phi }\rangle )^{2}$.

In the ideal multi-PSSs, the quantum state is given by $\left \vert \psi
_{\phi }\right \rangle =A\hat{U}_{\phi }\hat{a}^{m}\hat{b}^{n}\hat{B}%
_{1}\left \vert \psi \right \rangle _{in}$.

According to Eq. (\ref{s7}) and Eq. (\ref{s5}), the analytical expression of
the QFI can be derived as follows:%
\begin{eqnarray}
F &=&4[A^{2}D_{m+2,m+2,n,n}e^{Q}  \notag \\
&&+A^{2}D_{m+1,m+1,n,n}e^{Q}  \notag \\
&&-\left( A^{2}D_{m+1,m+1,n,n}e^{Q}\right) ^{2}].  \label{d1}
\end{eqnarray}

It is possible to explore the connections between the QFI and the related
parameters using in Eq. (\ref{d1}).
\begin{figure}[tbp]
\label{Fig10} {\centering \includegraphics[width=0.95\columnwidth]{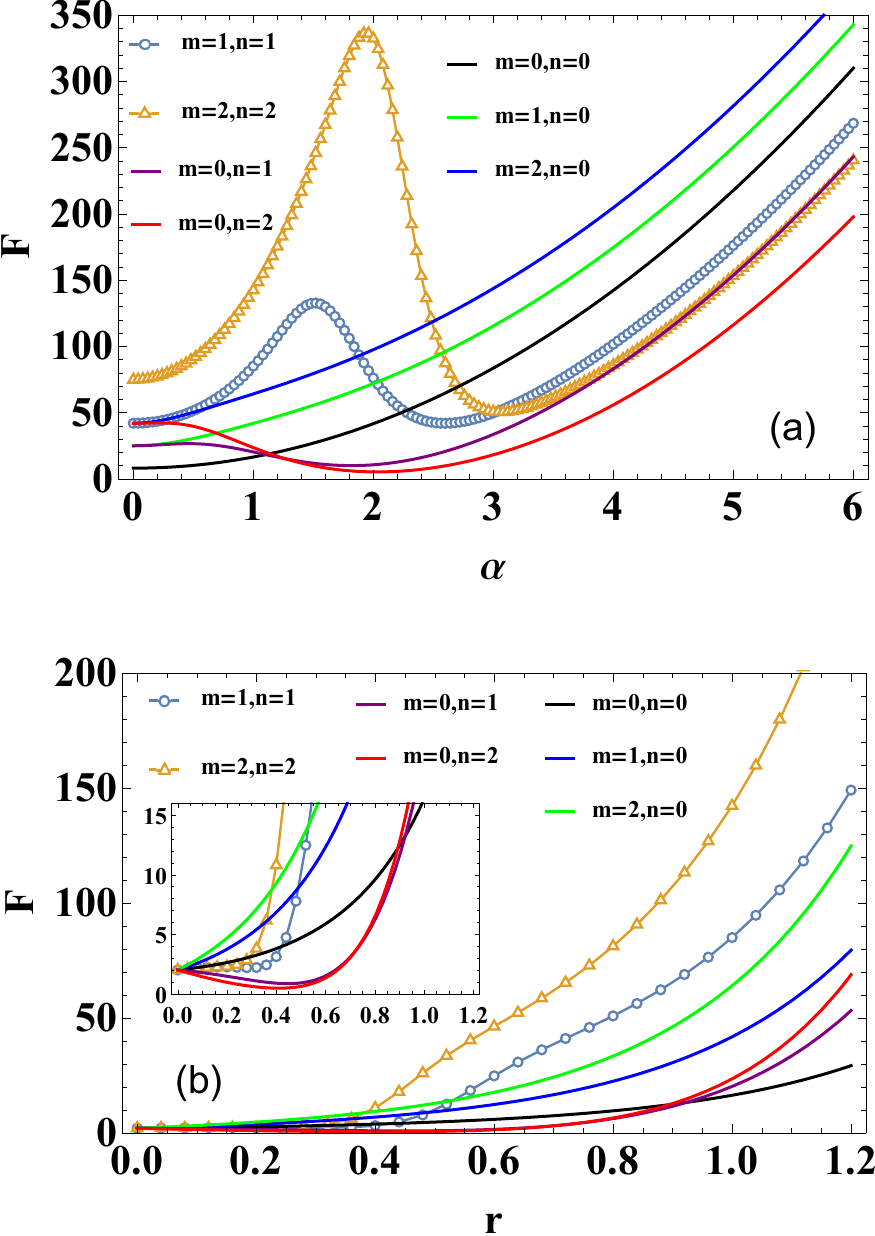}%
}
\caption{The QFI as a function of (a) the coherent amplitude $\protect \alpha
$, with $r=1$; (b) the squeezing parameter $r$, with $\protect \alpha =1$.}
\label{10}
\end{figure}

In the ideal case, we systematically analyze the effects of three mulit-PSSs
on the QFI. To see the effects of $m$ and $n$ on the QFI, the specific
parameters are set as follows: scheme A ($m=1,n=0$ and $m=2,n=0$), scheme B (%
$m=0,n=1$ and $m=0,n=2$), and scheme C ($m=1,n=1$ and $m=2,n=2$). The QFI as
a function of coherence amplitude and squeezing parameters are plotted, in
Fig. 10.

As shown in Figures 10(a) and 10(b), it can be observed that: (i) With fixed
$r=1$, the QFI of scheme A is similar to the standard QFI. Both increase
monotonically with $\alpha $ and remain consistently higher than the
standard, demonstrating significant improvement across the entire range; the
QFI of scheme B first decreases and then increases, with an improvement
effect on QFI only in a relatively small range; the QFI of scheme C shows a
non-monotonic change characteristic of first increasing, then decreasing,
and then increasing again. Scheme C significantly enhances the QFI over a
broader range of parameters and even\ outperforms scheme A within specific
parameter ranges. Especially, it is approximately around $\alpha <1.8$ ($%
r>0.5$), the QFI for scheme C ($m=1,n=1$) is higher than that for scheme A ($%
m=2,n=0$) and scheme B ($m=0,n=2$). (ii) When $\alpha $ is fixed at 1, the
QFI of scheme A monotonically increases with the increase of the squeezing
parameter $r$; the QFI of schemes B and C first slightly decreases and then
gradually increases. Among them, scheme C has the best improvement effect,
followed by scheme A, while scheme B performs poorly and only shows a
certain improvement effect on QFI when $r$ is relatively large.

Overall, scheme A and C have shown significant improvement effects on QFI,
especially demonstrating their respective advantages under different
parameter conditions, while the improvement range of scheme B is limited,
its effect is still closely related to the selection of relevant parameters.

Actually, the QFI can be related with the phase sensitivity via \cite{f01}%
\begin{equation}
\Delta \phi _{QCRB}=\frac{1}{\sqrt{vF}},  \label{d2}
\end{equation}%
where $v$ is the number of measurements. For simplicity, we set $v=1$. $%
\Delta \phi _{QCRB}$ is another quantum theoretical limit which does not
depend on a specific detection method \cite{f1,f2}.
\begin{figure}[tbp]
\label{Fig11} {\centering \includegraphics[width=0.95\columnwidth]{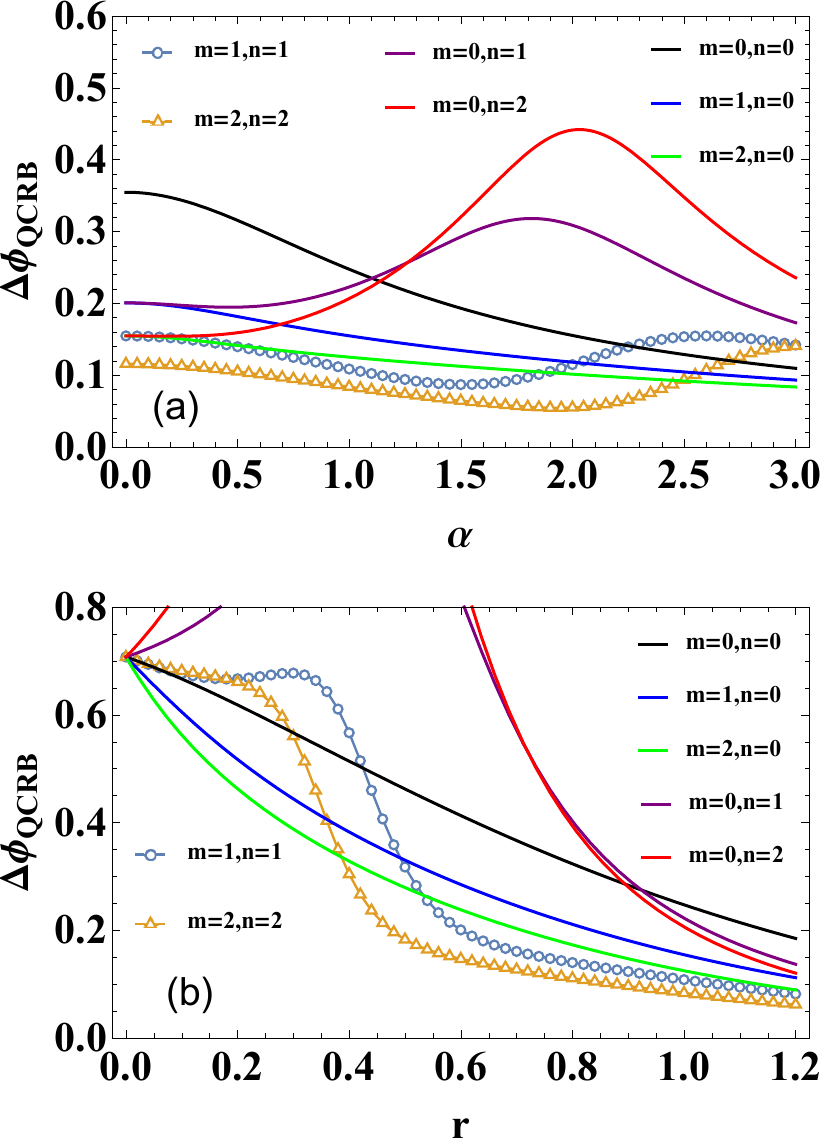}%
}
\caption{The $\Delta \protect \phi _{QCRB}$ as a function of (a) the coherent
amplitude $\protect \alpha $ with $r=1$; (b) the squeezing parameter $r$,
with $\protect \alpha =1$.}
\label{11}
\end{figure}

Fig. 11 shows $\Delta \phi _{QCRB}$ as a function of $\alpha $ $(r)$ for
given $r$ $(\alpha )$. In scheme A, $\Delta \phi _{QCRB}$ improves with
increasing $\alpha $ ($r$), becoming more significant as $m$ increases. In
scheme B, $\Delta \phi _{QCRB}$ only improves for small $\alpha $ (large $r$%
), with less effectiveness than scheme A. scheme C enhances $\Delta \phi
_{QCRB}$ over a broader range compared to scheme B and outperforms scheme A
when $\alpha $ is small and $r$ is not low.

\subsection{Photon losses case}

In this subsection, we extend our analysis to evaluate the QFI in the
presence of photon losses. In our scheme, the phase shift occurs on the $a$%
-path inside the MZI. For simplicity, we only consider the photon losses of
the mode $a$ , which can be modeled by fictitious BSs, as illustrated in
Fig. 12.
\begin{figure}[tbp]
\label{Fig12} {\centering \includegraphics[width=0.95\columnwidth]{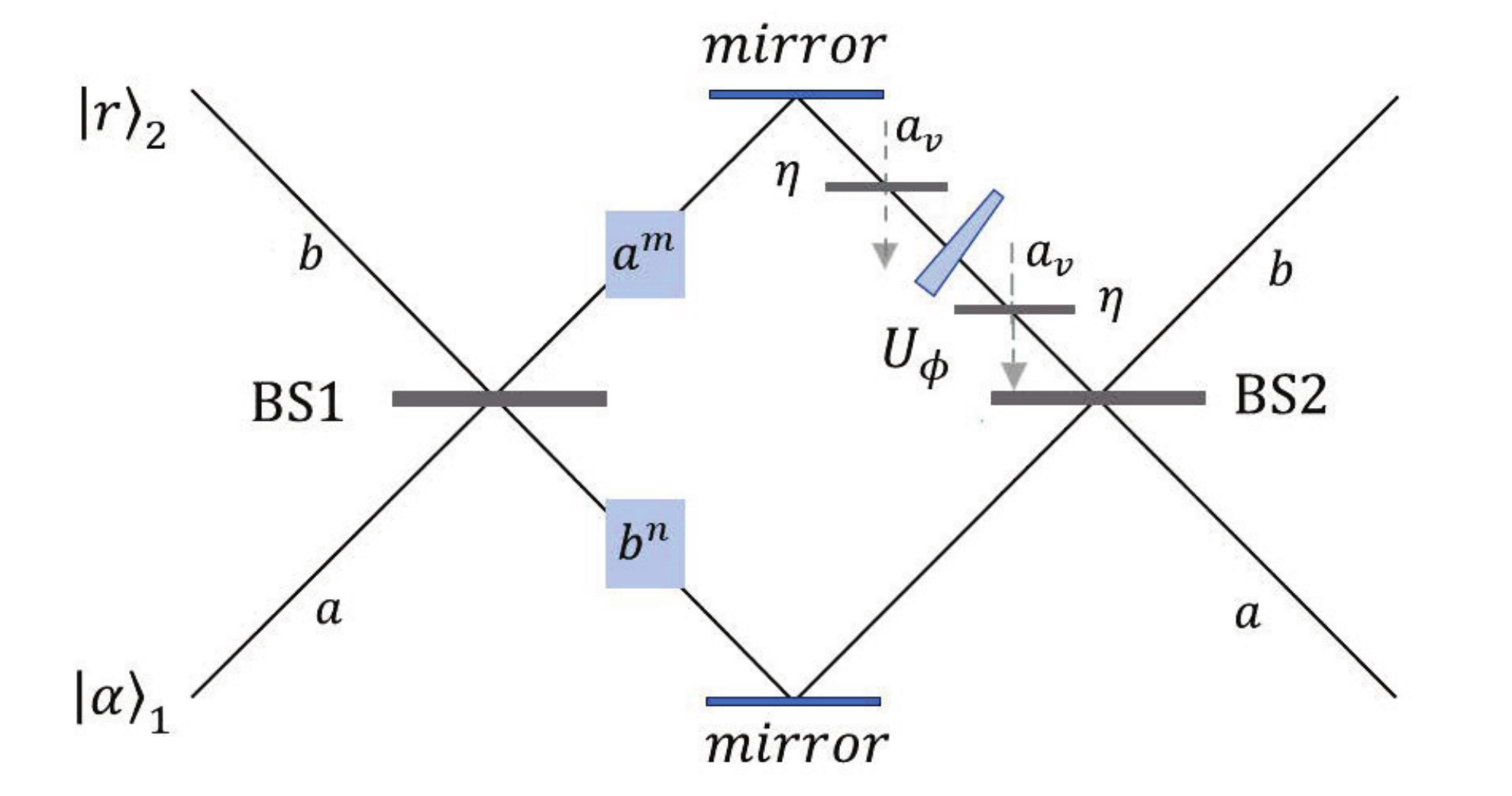}%
}
\caption{Schematic diagram of the photon losses on mode $a$. The losses
occurs before the PS operations.}
\label{12}
\end{figure}

For realistic quantum systems, it is difficult to calculate the QFI with
internal non-Gaussian operations directly. In the presence of photon losses,
we utilize the idea of purification limit, treating the quantum system S and
the environment E as an expanded isolated composite system, thereby
transforming the lossy evolution into a unitary evolution. When considering
photon losses, Kraus operators can be introduced, which could simplify the
calculation process. Next, we define the Kraus operator%
\begin{equation}
\hat{\Pi}_{l}=\sqrt{\frac{\left( 1-\eta \right) ^{l}}{l!}}\eta ^{\frac{\hat{n%
}_{a}}{2}}\hat{a}^{l}.  \label{dd3}
\end{equation}

Considering the two extreme cases of loss before and after the phase
shifter, the Kraus operator can be expressed as%
\begin{equation}
\hat{\Pi}_{l}\left( \phi ,\eta ,\gamma \right) =\sqrt{\frac{\left( 1-\eta
\right) ^{l}}{l!}}e^{i\phi \left( \hat{n}_{a}+\gamma l\right) }\eta ^{\frac{%
\hat{n}_{a}}{2}}\hat{a}^{l},  \label{d4}
\end{equation}%
and it satisfies%
\begin{equation}
\sum_{l}\hat{\Pi}_{l}^{\dagger }\left( \phi ,\eta ,\gamma \right) \hat{\Pi}%
_{l}\left( \phi ,\eta ,\gamma \right) =1.  \label{d5}
\end{equation}

In quantum systems, the transmittance $\eta $ of a virtual BS can be used to
describe the loss characteristics of the arm, $\eta =0$ and $\eta =1$
correspond to complete absorption and lossless conditions, respectively. $%
\gamma $ is the loss factor, with $\gamma =0$ and $\gamma =1$ corresponding
to the losses before and after the phase shifter respectively. According to
the method proposed by Escher \textit{et al}., considering the loss, the QFI
can be calculated as \cite{f3}:%
\begin{equation}
F_{L}=\min \limits_{\left \{ \hat{\Pi}_{l}\left( \phi ,\eta ,\lambda \right)
\right \} }C_{Q}\left[ \left \vert \psi _{s}\right \rangle \left \langle
\psi _{s}\right \vert ,\hat{\Pi}_{l}\left( \phi ,\eta ,\lambda \right) %
\right] .  \label{d6}
\end{equation}%
Here, $C_{Q}$ represents the QFI in the extended noise system, $\left \vert
\psi _{s}\right \rangle $ is the initial state of the detection system S. $%
\hat{\Pi}_{l}\left( \phi ,\eta ,\lambda \right) $ are Kraus operators, used
to describe the loss process of system S. Additionally, the mathematical
expression of $C_{Q}$ is as follows:%
\begin{equation*}
C_{Q}=4\left[ \left \langle \psi _{s}\right \vert \hat{H}_{1}^{2}\left \vert
\psi _{s}\right \rangle -\left \vert \left \langle \psi _{s}\right \vert
\hat{H}_{2}\left \vert \psi _{s}\right \rangle \right \vert ^{2}\right] ,
\end{equation*}%
where%
\begin{equation}
\hat{H}_{1}=\sum_{l}\frac{d\hat{\Pi}^{\dagger }\left( \phi ,\eta ,\lambda
\right) }{d\phi }\frac{d\hat{\Pi}\left( \phi ,\eta ,\lambda \right) }{d\phi }%
,  \label{d7}
\end{equation}%
and%
\begin{equation}
\hat{H}_{2}=i\sum_{l}\frac{d\hat{\Pi}^{\dagger }\left( \phi ,\eta ,\lambda
\right) }{d\phi }\hat{\Pi}\left( \phi ,\eta ,\lambda \right) .  \label{d8}
\end{equation}

By optimizing to minimize, the final expression of the QFI under photon
losses is \cite{f4}:%
\begin{equation}
F_{L}=\frac{4\eta \left \langle \hat{n}_{a}\right \rangle F}{\left( 1-\eta
\right) F+4\eta \left \langle \hat{n}_{a}\right \rangle },  \label{d9}
\end{equation}%
where $F$ is the QFI in the ideal case, $\eta $ is the\ transmittance, and $%
\left \langle \hat{n}_{a}\right \rangle $ is the total average photon number
of mode $a$ within the MZI. Hence, according to Eqs. (\ref{s7}) and (\ref{d9}%
) the expression of the QFI in the presence of photon losses is as follows:%
\begin{equation}
F_{L}=\frac{4\eta \left( A^{2}D_{m+1,m+1,n,n}e^{Q}\right) F}{\left( 1-\eta
\right) F+4\eta \left( A^{2}D_{m+1,m+1,n,n}e^{Q}\right) }.  \label{d10}
\end{equation}

Under photon losses conditions, we analyze the impact of various parameters
on QFI to characterize its degradation. As shown in Figure 13(a), with fixed
parameters $\alpha =1$, $r=1$: (i) QFI increases with higher transmittance $%
\eta $; (ii) Both scheme A and C significantly improve QFI under photon\
losses, with improvements increasing as $m$ and $n$ grow. And scheme C
exhibits a higher QFI than scheme A; (iii) Scheme B only slightly improves
QFI at low loss levels and performs poorly overall.
\begin{figure}[tbp]
\label{Fig13} {\centering \includegraphics[width=0.95\columnwidth]{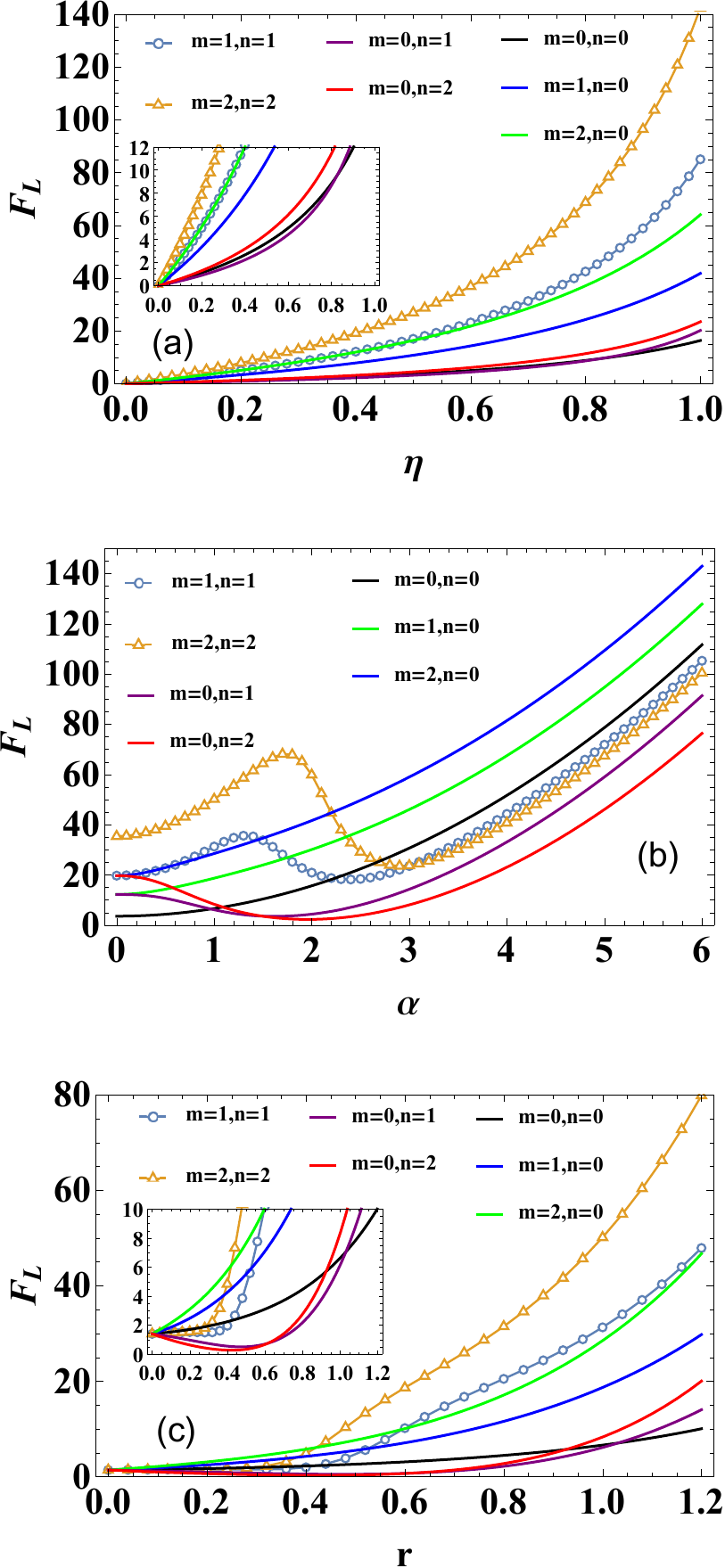}%
}
\caption{The $F_{L}$ as a function of (a) transmittance $\protect \eta $,
with $\protect \alpha =1$ and $r=1$;\ (b) the coherent amplitude $\protect%
\alpha $, with $r=1$ and $\protect \eta =0.8$; (c) the squeezing parameter $r$%
, with $\protect \alpha =1$\ and $\protect \eta =0.8$.}
\label{13}
\end{figure}

In Figures 13(b) and (c), with fixed the transmittance $\eta =0.7$, the
variation trend of QFI with other parameters under losses conditions closely
resembles that in the ideal scenario. Specifically, as shown in Figure
13(b), the QFI of scheme A is similar to the standard QFI, both
monotonically increase with the increase of the coherent amplitude; the QFI
of scheme B first decreases and then increases, while that of scheme C first
increases, then decreases and increases again. From this trend, it can be
seen that the maximum value of QFI will be achieved when the coherent
amplitude is relatively large. Within the range of large coherent amplitude,
only the QFI of scheme A is consistently higher than the standard scheme,
showing significant improvement. However, within the range of small coherent
amplitude, the QFI values of schemes B and C are higher than the standard
situation in some cases, still demonstrating certain improvement
capabilities. Notably, under small coherent amplitude, the QFI of scheme C
is superior to that of scheme A. And then, the QFI of scheme A increases
with increasing of the squeezing parameter $r$, but for schemes B and C, it
initially decreases slightly before increasing. Overall, scheme C performs
optimally across most parameter ranges, whereas scheme B is less effective
than the other two schemes but still shows improvement within a larger
squeezing parameter range. It is that the multi-PSSs alters the trend of QFI
with the change of parameters. In summary, with fixed parameters $\alpha
=1,r=1$ under photon losses, scheme C is the optimal scheme, followed by
scheme A, while scheme B performs poorly.

Similar to the ideal case, one can compute the QCRB as $\Delta \phi
_{QCRBL}=1/\sqrt{vF_{L}}$\ and for simplicity we take $v=1$. As shown in
Figure 14, $\Delta \phi _{QCRBL}$ decreases as the transmittance $\eta $
decreases.\ And could be furtherly improved with the increase of the
photon-subtracted number $m$ or $n$. Scheme C is the best, followed by
scheme A. Besides, the $\Delta \phi _{QCRBL}$ varies with the coherent
amplitude and squeezing parameter similarly to the ideal case.
\begin{figure}[tbp]
\label{Fig14} {\centering \includegraphics[width=0.95\columnwidth]{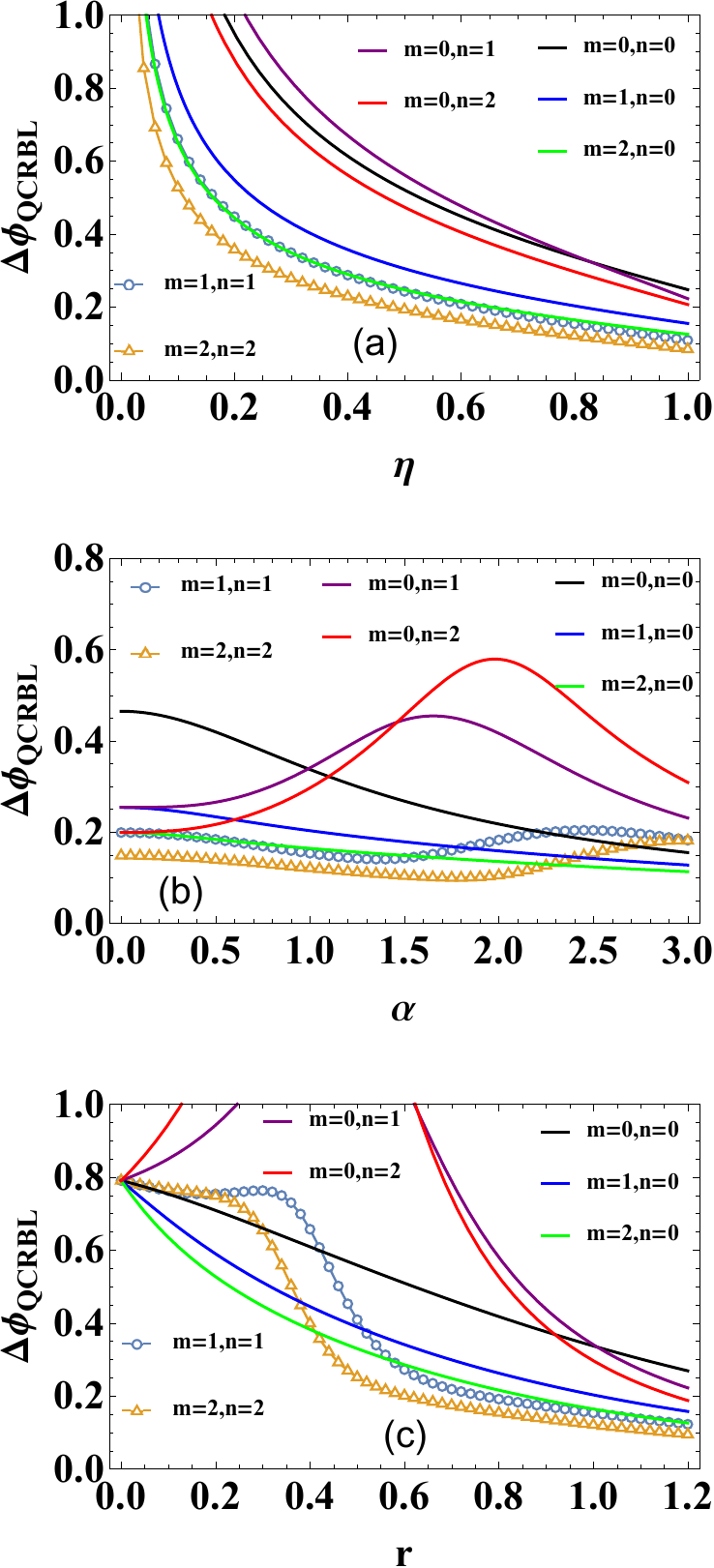}%
}
\caption{The $\Delta \protect \phi _{QCRBL}$ as a function of (a)
transmittance $\protect \eta $, with $\protect \alpha =1$ and $r=1$,\ (b) the
coherent amplitude $\protect \alpha $ with $r=1$ and $\protect \eta =0.8$ (c)
the squeezing parameter $r$, with $\protect \alpha =1$\ and $\protect \eta %
=0.8 $.}
\label{14}
\end{figure}

\section{Conclusion}

To enhance phase sensitivity, we propose multi-PSSs and investigate their
phase sensitivities under different detection methods. Based on various
intensity and homodyne detection methods, we performed the same
photon-subtracted number for all three schemes. The results show that: (i)
Intensity difference detection is the optimal intensity detection method,
with scheme C providing the best improvement in phase sensitivity; schemes A
and B have identical effects. (ii) $X_{b}$ is the optimal homodyne detection
method, but only scheme A improves phase sensitivity under this method,
while schemes B and C fail to improve it and even weaken it. These findings
indicate that the choice of detection method significantly impacts phase
sensitivity.

Secondly, based on these two finalized optimal detection methods, i.e.,
intensity difference detection and mode-$b$ homodyne detection, we further
analyze the effect of multi-PSSs on phase sensitivity. By adopting scheme A,
we study the variation of phase sensitivity with coherent amplitude,
squeezing parameter, and transmittance, and compare it with theoretical
limits. Under intensity difference detection, the phase sensitivity of
scheme A performs better at smaller coherent amplitudes and larger squeezing
parameters. When $r=1$ and $\alpha =1$, in the ideal case, subtracting two
photons from scheme A ($m=2$) can break the SQL, and subtracting three
photons ($m=3$) can surpass the HL. In the lossy case, scheme A still breaks
the SQL and approaches the HL at $m=3$. For $X_{b}$ homodyne detection,
scheme A's phase sensitivity increases with the increase of coherent
amplitude and squeezing parameter. In the ideal case, the standard MZI
scheme breaks the HL within a limited range, while scheme A surpasses the HL
over a broader range, making a further improvement with increasing the
photon-subtracted number $m$. In the lossy case, the standard scheme cannot
break the SQL, but our scheme still surpasses the HL, even at $m=1$. From
the above, scheme A can significantly enhance the phase sensitivity and
restrain the internal losses in the MZI. Additionally, we study the
influence of three multi-PSSs on QFI, and compare them under the same
parameters. The results show that within certain parameter ranges, scheme C
performed the best, followed by scheme A, while scheme B had limited
improvement.

In conclusion, the multi-PSSs effectively enhance the quantum measurement
precision of the MZI and overcome internal photon losses. This research
demonstrates the potential of photon subtraction operations in improving the
performance\ of\ quantum metrology and information processing systems.

\begin{acknowledgments}
\bigskip This work is supported by the National Natural Science Foundation
of China (Grants No.11964013 and No. 12104195) and the Jiangxi Provincial
Natural Science Foundation (Grants No. 20242BAB26009 and 20232BAB211033),
Jiangxi Provincial Key Laboratory of Advanced Electronic Materials and
Devices (Grant No. 2024SSY03011), as well as Jiangxi Civil-Military
Integration Research Institute (Grant No. 2024JXRH0Y07).
\end{acknowledgments}

\textbf{APPENDIX\ A : THE PHASE SENSITIVITY} \bigskip

In this Appendix, we first provide the calculation formulas of the phase
sensitivity based on intensity detection for the multi-PSSs as follows%
\begin{equation}
\Delta \phi _{1}=\frac{\sqrt{c_{1}^{2}\left \langle \Delta ^{2}\hat{N}%
_{a}\right \rangle +d_{1}^{2}\left \langle \Delta ^{2}\hat{N}_{b}\right
\rangle +2c_{1}d_{1}cov\left[ \hat{N}_{a},\hat{N}_{b}\right] }}{\left \vert
\partial _{\phi }\left( c_{1}\left \langle \hat{N}_{a}\right \rangle
+d_{1}\left \langle \hat{N}_{b}\right \rangle \right) \right \vert },
\tag{A1}
\end{equation}%
where $\left \langle \Delta ^{2}\hat{N}_{a}\right \rangle =\left \langle
\hat{N}_{a}^{2}\right \rangle -\left \langle \hat{N}_{a}\right \rangle
^{2},\left \langle \Delta ^{2}\hat{N}_{b}\right \rangle =\left \langle \hat{N%
}_{b}^{2}\right \rangle -\left \langle \hat{N}_{b}\right \rangle ^{2},cov%
\left[ \hat{N}_{a},\hat{N}_{b}\right] =\left \langle \hat{N}_{a}\hat{N}%
_{b}\right \rangle -\left \langle \hat{N}_{a}\right \rangle \left \langle
\hat{N}_{b}\right \rangle .$

And we provide the calculation formulas of the phase sensitivity based on
homodyne detection for the multi-PSSs as follows%
\begin{equation}
\Delta \phi _{2}=\frac{\sqrt{c_{2}^{2}\left \langle \Delta ^{2}\hat{X}%
_{a}\right \rangle +d_{2}^{2}\left \langle \Delta ^{2}\hat{X}_{b}\right
\rangle +2c_{2}d_{2}cov\left[ \hat{X}_{a},\hat{X}_{b}\right] }}{\left \vert
\partial _{\phi }\left( (c_{2}\left \langle \hat{X}_{a}\right \rangle
+d_{2}\left \langle \hat{X}_{b}\right \rangle )\right) \right \vert },
\tag{A2}
\end{equation}%
where $\left \langle \Delta ^{2}\hat{X}_{a}\right \rangle =\left \langle
\hat{X}_{a}^{2}\right \rangle -\left \langle \hat{X}_{a}\right \rangle
^{2},\  \left \langle \Delta ^{2}\hat{X}_{b}\right \rangle =\left \langle
\hat{X}_{b}^{2}\right \rangle -\left \langle \hat{X}_{b}\right \rangle
^{2},\ cov[\hat{X}_{a},\hat{X}_{b}]=\left \langle \hat{X}_{a}\hat{X}%
_{b}\right \rangle -\left \langle \hat{X}_{a}\right \rangle \left \langle
\hat{X}_{b}\right \rangle .$

Thus, we need to calculate the expected mechanical quantities related to the
phase sensitivity of multi-PSSs. For the convenience of calculation, we have
calculated the general formula for the expected value of the output end,
i.e. $\left \langle \hat{a}^{\dagger p_{1}}\hat{a}^{p_{2}}\hat{b}^{\dagger
q_{1}}\hat{b}^{q_{2}}\right \rangle $. In our paper, the output state $%
\left
\vert \Psi \right \rangle _{out}$ is given by Eq. (\ref{b0}).
Substituting Eq. (\ref{b0}) into the general formula, we can obtain its
expression given by Eq. (\ref{b1}). The normalization constant for the
multi-PSSs, denoted by $A$, is given by Eq. (\ref{06}).

According to Eqs. (\ref{b1}), so the expectations related to the phase
sensitivity based on intensity detection for multi-PSSs are specifically
calculated as%
\begin{align}
\left \langle \Delta ^{2}\hat{N}_{a}\right \rangle & =[A^{2}\left(
D_{m,n,2,2,0,0}+D_{m,n,1,1,0,0}\right) e^{M}  \notag \\
& -\left( A^{2}D_{m,n,1,1,0,0}e^{M}\right) ^{2}],  \tag{A3}
\end{align}%
and%
\begin{align}
\left \langle \Delta ^{2}\hat{N}_{b}\right \rangle & =[A^{2}\left(
D_{m,n,0,0,2,2}+D_{m,n,0,0,1,1}\right) e^{M}  \notag \\
& -\left( A^{2}D_{m,n,0,0,1,1}e^{M}\right) ^{2}],  \tag{A4}
\end{align}%
and%
\begin{align}
cov\left[ \hat{N}_{a},\hat{N}_{b}\right] & =(A^{2}D_{m,n,1,1,1,1}e^{M}
\notag \\
& -A^{2}D_{m,n,0,0,1,1}e^{M}  \notag \\
& \times A^{2}D_{m,n1,1,0,0,}e^{M}).  \tag{A5}
\end{align}%
And according to Eqs. (\ref{b1}), so the expectations related to the phase
sensitivity based on homodyne detection for multi-PSSs are specifically
calculated as%
\begin{align}
\left \langle \Delta ^{2}\hat{X}_{a}\right \rangle &
=[A^{2}(D_{m,n,2,0,0,0}+D_{m,n,0,2,0,0}  \notag \\
& +2D_{m,n,1,1,0,0})e^{M}+1  \notag \\
& -(A^{2}(D_{m,n,1,0,0,0}  \notag \\
& +D_{m,n,0,1,0,0})e^{M})^{2}],  \tag{A6}
\end{align}%
and%
\begin{align}
\left \langle \Delta ^{2}\hat{X}_{b}\right \rangle &
=[A^{2}(D_{m,n,0,0,2,0}+D_{m,n,0,0,0,2}  \notag \\
& +2D_{m,n,0,0,1,1})e^{M}+1  \notag \\
& -(A^{2}(D_{m,n,0,0,1,0}  \notag \\
& +D_{m,n,0,0,0,1})e^{M})^{2}],  \tag{A7}
\end{align}%
and%
\begin{align}
cov\left[ \hat{X}_{a},\hat{X}_{b}\right] & =(A^{2}D_{m,n,1,1,1,1}e^{M}
\notag \\
& -A^{2}D_{m,n,0,0,1,1}e^{M}  \notag \\
& \times A^{2}D_{m,n1,1,0,0,}e^{M}).  \tag{A8}
\end{align}


\begin{thebibliography}{99}
\bibitem{02} Kullback S. Information theory and statistics[M]. Courier
Corporation (1997).

\bibitem{03} Luis A. Equivalence between macroscopic quantum superpositions
and maximally entangled states: Application to phase-shift detection[J].
Physical Review A, 64(5):54102 (2001).

\bibitem{04} Pezze L, Smerzi A. Phase Sensitivity of a Mach-Zehnder
Interferometer[J]. Physical Review A, 73(1):5689-5693 (2005).

\bibitem{05} Uys H, Meystre P. Quantum states for Heisenberg-limited
interferometry[J]. Physical Review A, 76(1):013804 (2007).

\bibitem{07} LIGO Scientific Collaboration. A gravitational wave observatory
operating beyond the quantum shot-noise limit. Nat Phys,7(12):962--965
(2011).

\bibitem{3} Boto A N, Kok P, Abrams D S, et al. Quantum Interferometric
Optical Lithography: Exploiting Entanglement to Beat the Diffraction Limit
[J]. Physical Review Letters, 85(13): 2733-6 (2000).

\bibitem{4} Fonseca E J S, Monken C H,\ P\'{a}dua S. Measurement of the de
Broglie Wavelength of a Multiphoton Wave Packet [J]. Physical Review
Letters, 82(14): 2868-71 (1999).

\bibitem{5} Tsang M. Quantum imaging beyond the diffraction limit by optical
centroid measurements[J]. Physical Review Letters, 102(25): 253601 (2009).

\bibitem{6} Shin H, Chan K W C, Chang H J, et al. Quantum spatial
superresolution by optical centroid measurements[J]. Physical Review
Letters, 107(8): 083603 (2011).

\bibitem{7} Ludlow A D, Boyd M M, Ye J, et al. Optical atomic clocks [J].
Reviews of Modern Physics, 87(2): 637-701 (2015).

\bibitem{8} Diddams S A, Bergquist J C, Jefferts S R, et al. Standards of
Time and Frequency at the Outset of the 21st Century [J]. Science,
306(5700): 1318-24 (2004).

\bibitem{9} Abbott B P, et al. GW150914: Implications for the Stochastic
Gravitational-Wave Background from Binary Black Holes[J]. Physical Review
Letters,116(13):131102 (2016).

\bibitem{10} Lasky P D, Thrane E, Levin Y, et al. Detecting
Gravitational-Wave Memory with LIGO: Implications of GW150914[J]. Physical
Review Letters, 117(6):061102 (2016).

\bibitem{12} Durfee, D. S., Shaham, Y. K., \& Kasevich, M. A. Long-term
stability of an area-reversible atom-interferometer Sagnac gyroscope.
Physical review letters, 97(24), 240801 (2006).

\bibitem{13} Dowling J P. Correlated input-port, matter-wave interferometer:
Quantum-noise limits to the atom-laser gyroscope[J]. Physical Review A,
57(6):4736-4746 (1998).

\bibitem{15} Giovannetti V, Lloyd S, Maccone L. Quantum Metrology[J].
Physical Review Letters, 96(1):010401 (2006).

\bibitem{a1} Holland M J, Burnett K. Interferometric detection of optical
phase shifts at the Heisenberg limit[J]. Physical Review Letters, 71(9):
1355 (1993).

\bibitem{a2} Giovannetti V, Lloyd S, Maccone L. Quantum-enhanced
measurements: beating the standard quantum limit[J]. Science, 306(5700):
1330-1336 (2004).

\bibitem{a3} T\'{o}th G, Apellaniz I. Quantum metrology from a quantum
information science perspective[J]. Journal of Physics A: Mathematical and
Theoretical, 47(42): 424006 (2014).

\bibitem{a4} D'Ariano G M, Macchiavello C, Sacchi M F. On the general
problem of quantum phase estimation [J]. Physics Letters A, 248(2): 103-8
(1998).

\bibitem{c1} Caves C M. Quantum-mechanical noise in an interferometer[J].
Physical Review D, 23(8): 1693 (1981).

\bibitem{c3} Joo J, Munro W J, and Spiller T P. Quantum Metrology with
Entangled Coherent States[J]. Physical Review Letters, 107(8): p.94-97
(2011).

\bibitem{c6} Dowling J P. Quantum optical metrology? - the lowdown on
high-N00N states[J]. Contemporary Physics, 49(2):125-143 (2008).

\bibitem{c7} Rarity J R, Tapster P R, Jakeman E, Larchuk T, Campos R A,
Teich M C, and Saleh B E A. Two-photon interference in a Mach-Zehnder
interferometer[J]. Physical Review Letters,65(11):1348-1351 (1900).

\bibitem{c8} Anisimov P M, Raterman G M, Chiruvelli A, et al. Quantum
Metrology with Two-Mode Squeezed Vacuum: Parity Detection Beats the
Heisenberg Limit[J]. Physical Review Letters, 104(10): p.103602.1-103602.4
(2010).

\bibitem{c10} Yurke B, Mccall S L, Klauder J R. SU(2) and SU(1,1)
interferometers[J]. Physical Review A, 33(6):4033-4054 (1986).

\bibitem{c101} Wei, C. P., \& Zhang, Z. M. Improving the phase sensitivity
of a Mach--Zehnder interferometer via a nonlinear phase shifter. Journal of
Modern Optics, 64(7), 743-749 (2017).

\bibitem{c102} Jiao, G. F., Zhang, K., Chen, L. Q., Zhang, W., \& Yuan, C.
H. Nonlinear phase estimation enhanced by an actively correlated
Mach-Zehnder interferometer. Physical Review A, 102(3), 033520 (2020).

\bibitem{c103} Huang, W., Liang, X., Yuan, C. H., Zhang, W., \& Chen, L. Q.
Optimal phase measurements in a lossy Mach--Zehnder interferometer with
coherent input light. Results in Physics, 50,106574 (2023).

\bibitem{c104} Pan, P., Wen, J., Zha, S., Cai, X., Ma, H., \& An, J.
Fabrication and error analysis of a InGaAsP/InP polarization beam splitter
based on an asymmetric Mach-Zehnder interferometer. Optical Materials, 118,
111250 (2021).

\bibitem{e1} Kumar, C., Rishabh, Sharma, M., \& Arora, S.
Parity-detection-based Mach-Zehnder interferometry with coherent and
non-Gaussian squeezed vacuum states as inputs. Physical Review A, 108(1),
012605 (2023).

\bibitem{e2} Tang, X. B., Gao, F., Wang, Y. X., Kuang, S., \& Shuang, F.
Non-Gaussian quantum states generation and robust quantum non-Gaussianity
via squeezing field. Chinese Physics B, 24(3), 034208 (2015).

\bibitem{e3} Chen, X. Y. The entanglement properties of non-Gaussian states
prepared by photon subtraction from two-mode squeezed thermal states.
Physics Letters A, 372(17), 2976-2979 (2008).

\bibitem{e4} Su, D., Myers, C. R., \& Sabapathy, K. K. Conversion of
Gaussian states to non-Gaussian states using photon-number-resolving
detectors. Physical Review A, 100(5), 052301 (2019).

\bibitem{e6} Kumar, C., \& Arora, S. Success probability and performance
optimization in non-Gaussian continuous-variable quantum teleportation.
Physical Review A, 107(1), 012418 (2023).

\bibitem{e7} Chuong, H. S., \& Duc, T. M. Enhancement of non-Gaussianity and
nonclassicality of pair coherent states by superposition of photon addition
and subtraction. Journal of Physics B: Atomic, Molecular and Optical
Physics, 56(20), 205401 (2023).

\bibitem{e8} Zhang, X. Y., Yang, C. Y., Wang, J. S., \& Meng, X. G.
Non-Gaussian quantum states generated via quantum catalysis and their
statistical properties. Chinese Physics B, 33(4), 040308 (2024).

\bibitem{e10} Gagatsos, C. N., \& Guha, S. Efficient representation of
Gaussian states for multimode non-Gaussian quantum state engineering via
subtraction of arbitrary number of photons. Physical Review A, 99(5), 053816
(2019).

\bibitem{e11} Kumar, C., Sharma, M., \& Arora, S. Continuous Variable
Quantum Teleportation in a Dissipative Environment: Comparison of
Non-Gaussian Operations Before and After Noisy Channel. Advanced Quantum
Technologies, 7(4), 2300344 (2024).

\bibitem{d01} Zavatta,A.,Parigi,V.,Kim,M.S.,and.Bellini,M.,Subtracting
photons from.arbitrary lightfields: Experimental test of coherent state
invariance by single-photon annihilation[J].New Journal of
Physics,10(12):123006 (2008).

\bibitem{d02} Wenger,J.,Tualle-Brouri,R.,and Grangier,P.,Non-Gaussian
statistics from individual pulses of squeezed light[J].Physical Review
Letters, 92(15):153601 (2004).

\bibitem{d03} Zhang, S. High-probability photon catalysis with non-linear
photon beam splitter. Optik, 276, 170638 (2023).

\bibitem{d21} Verma, M., Kumar, C., Mishra, K. K., \& Panigrahi, P. K.
Advantage of Non-Gaussian Operations in Phase Estimation via Mach--Zehnder
Interferometer. Advanced Quantum Technologies, 2400192 (2024).

\bibitem{d22} Kumar, C., Rishabh, \& Arora, S. Enhanced phase estimation in
parity-detection-based Mach--Zehnder interferometer using non-Gaussian
two-mode squeezed thermal input state. Annalen der Physik, 535(8), 2300117
(2023).

\bibitem{d23} Y. K. Xu, T. Zhao, Q. Q. Kang, C. J. Liu, L. Y. Hu, and S. Q.
Liu, Phase sensitivity of an SU(1,1) interferometer in photon-loss via
photon operations, Opt. Express 31, 8414 (2023).

\bibitem{d24} Kang, Q., Zhao, Z., Xu, Y., Zhao, T., Liu, C., \& Hu, L. Phase
estimation via multi-photon subtraction inside the SU (1, 1) interferometer.
Physica Scripta, 99(8), 085111 (2024).

\bibitem{d25} Parks, A. D., Spence, S. E., Troupe, J. E., \& Rodecap, N. J.
Tripartite loss model for Mach-Zehnder interferometers with application to
phase sensitivity. Review of scientific instruments, 76(4) (2005).

\bibitem{d26} Wenger, J., Tualle-Brouri, R., \& Grangier, P. Non-Gaussian
statistics from individual pulses of squeezed light. Physical review
letters, 92(15), 153601 (2004).

\bibitem{d28} R. Demkowicz-Dobrza\'{n}ski, M. Jarzyna, and J.Ko\l ody\'{n}%
ski, Quantum limits in optical interferometry, Prog. Optics 60, 345-435
(2015).

\bibitem{d29} M. Bradshaw, P. K. Lam, and S. M. Assad, Ultimate precision of
joint quadrature parameter estimation with a Gaussian probe, Phys. Rev. A
97(1), 012106 (2018).

\bibitem{d3} Yuen H P, Chan V W S. Noise in Homodyne and Heterodyne
Detection[J]. Optics Letters, 8(3):177-179 (1983).

\bibitem{d5} Manceau M, Khalili F, Chekhova M. Improving the phase
super-sensitivity of squeezing-assisted interferometers by squeeze factor
unbalancing[J]. New Journal of Physics, 19(1):013014 (2017).

\bibitem{d7} Walls D. Squeezed states of light[J]. Nature, 306(5939):141-146
(1983).

\bibitem{d8} Szigeti S S, Lewis-Swan R J, Haine S A. Pumped-Up SU(1,1)
Interferometry[J]. Physical Review Letters, 118(15):150401 (2017).

\bibitem{d9} Ataman, S., Preda, A., \& Ionicioiu, R. Phase sensitivity of a
Mach-Zehnder interferometer with single-intensity and difference-intensity
detection. Physical Review A, 98(4), 043856 (2018).

\bibitem{d11} Bollinger J J, Itano W M, Wineland D J, et al. Optimal
frequency measurements with maximally correlated states[J]. Physical Review
A, 54(6):R4649-R4652 (1996).

\bibitem{d12} Anisimov P M, Raterman G M, Chiruvelli A, et al. Quantum
Metrology with Two-Mode Squeezed Vacuum: Parity Detection Beats the
Heisenberg Limit[J]. Physical review letters, 104(10):p.103602.1-103602.4
(2010).

\bibitem{d14} C. M. Caves, Quantum-mechanical noise in an interferometer,
Phys. Rev. D 23(8), 1693 (1981).

\bibitem{d15} O. Assaf and Y. Ben-Aryeh, Quantum mechanical noise in
coherent-state and squeezed-state Michelson interferometers, J. Opt. B:
Quantum Semiclass. Opt. 4(1), 49 (2002).

\bibitem{d16} J. Beltran and A. Luis, Breaking the Heisenberg limit with
inefficient detectors, Phys. Rev. A 72(4),045801 (2005).

\bibitem{f0} J. Liu, X. Jing, W. Zhong, and X. Wang, Quantum Fisher
Information for density matrices with arbitrary ranks, Commun. Theor. Phys.
61, 45-50 (2014).

\bibitem{f01} Helstrom C W, Quantum detection and estimation theory
(Academic), 123 (1976).

\bibitem{f1} C. W. Helstrom, Minimum mean-squared error of estimates in
quantum statistics, Phys. Lett. A 25(2), 101 (1967).

\bibitem{f2} C. W. Helstrom, Quantum detection and estimation theory, J.
Stat. Phys. 1(2), 231 (1969).

\bibitem{f3} B. M. Escher, R. L. de Matos Filho, and L. Davidovich, General
framework for estimating the ultimate precision limit in noisy
quantum-enhanced metrology, Nat. Phys. 7, 406 (2011).

\bibitem{f4} S. K. Chang, W. Ye, H. Zhang, L. Y. Hu, J. H. Huang, and S. Q.
Liu, Improvement of phase sensitivity in an SU(1, 1) interferometer via a
phase shift induced by a Kerr medium, Phys. Rev. A 105(3), 033704 (2022).
\end{thebibliography}
\end{document}